\documentclass[conference]{IEEEtran}
\IEEEoverridecommandlockouts

\usepackage{cite}
\usepackage{siunitx}
\usepackage{mathtools}
\usepackage{amsmath,amssymb,amsfonts}
\usepackage{algorithmic}
\usepackage{graphicx}
\usepackage{textcomp}
\usepackage{xcolor}
\usepackage[table]{xcolor}
\usepackage{tabularx}
\usepackage{makecell}
\usepackage{pifont}
\usepackage{booktabs}
\usepackage{threeparttable}
\usepackage[hyphens]{url}
\usepackage{multirow}

\def\BibTeX{{\rm B\kern-.05em{\sc i\kern-.025em b}\kern-.08em
    T\kern-.1667em\lower.7ex\hbox{E}\kern-.125emX}}
\begin{document}

\pdfpagewidth=8.5in
\pdfpageheight=11in

\newcommand{\iscasubmissionnumber}{NaN}
\newcommand\revise[1]{\textcolor{blue}{#1}}
\newcommand\postrebuttal[1]{\textcolor{red}{#1}}
\pagenumbering{arabic}

\title{HE\textsuperscript{2}: A Communication-Light Heterogeneous Architecture for Efficient Fully Homomorphic Encryption
\footnotesize \thanks{Corresponding author is Husheng Han.}}

\author{
    Shangyi Shi\textsuperscript{1,2,3}, Husheng Han\textsuperscript{1}, Zhaoxuan Kan\textsuperscript{1,2,3}, Yinghao Yang\textsuperscript{1}, Jianan Mu\textsuperscript{1},\\ Tenghui Hua\textsuperscript{1}, Ge Yu\textsuperscript{1,2,4}, Xinyao Zheng\textsuperscript{1,2,3}, Ling Liang\textsuperscript{5}, Zidong Du\textsuperscript{1,2}, Xing Hu\textsuperscript{1,2}\\
    \textsuperscript{1}\textit{State Key Laboratory of Processors, Institute of Computing Technology, CAS, Beijing, China} \\
    \textsuperscript{2}\textit{University of Chinese Academy of Sciences, Beijing, China} \quad \textsuperscript{3}\textit{Cambricon Technologies} \\
    \textsuperscript{4}\textit{School of Advanced Interdisciplinary Sciences, CAS, Beijing, China} \\
    \textsuperscript{5}\textit{School of Integrated Circuits, Peking University, Beijing, China} \\
    
    \{shishangyi22s, hanhusheng, kanzhaoxuan23z, yangyinghao\}@ict.ac.cn\\
    \{mujianan, huatenghui24s, yuge23s, zhengxinyao22s\}@ict.ac.cn\\ lingliang@pku.edu.cn \quad \{duzidong, huxing\}@ict.ac.cn \\
}

\maketitle
\thispagestyle{plain}
\pagestyle{plain}


\begin{abstract}
CKKS, an emerging fully homomorphic encryption (FHE) scheme, has been promising in privacy-preserving applications by enabling SIMD fixed-point computations on ciphertexts. Despite its strong security guarantees, CKKS involves both compute-intensive operators (ComOps) with high computational cost and memory-intensive operators (MemOps) with large memory footprints, making existing ASIC-based or NMP-based acceleration approaches suffer from high hardware overhead and limited efficiency. This observation motivates the integration of the architectural advantages of both paradigms into a heterogeneous xPU (ASIC)-xMU (NMP) architecture. However, in such a design, frequent and long-latency heterogeneous communication caused by the dominant keyswitch operator remains a key performance bottleneck.

In this paper, we propose HE\textsuperscript{2}, a communication-light xPU-xMU heterogeneous FHE accelerator with dataflow graph (DFG) optimization and architecture co-design.
First, we observe that the majority of communication arises at the interface between ModUp/ModDown and neighboring MemOps. To address this, we propose a DFG-level optimization framework to fully exploit the ModUp/ModDown reduction potential of the hoisting algorithm by identifying parallel keyswitch blocks and fusing them for reduced communication frequency. 
Second, we design an efficient heterogeneous architecture that adopts a group-level pipelined execution to effectively hide communication latency by leveraging the inherent parallelism across decomposed groups.
End-to-end evaluation results show that HE\textsuperscript{2} achieves 1.66$\times$ speedup and 9.23$\times$ lower EDAP (Energy-Delay-Area Product) compared to the state-of-the-art accelerator, with communication stalls accounting for only 6.67\% of the total latency. 
\end{abstract}
\section{Introduction}
\begin{figure}[]
    \centering
    \includegraphics[width=0.95\linewidth]{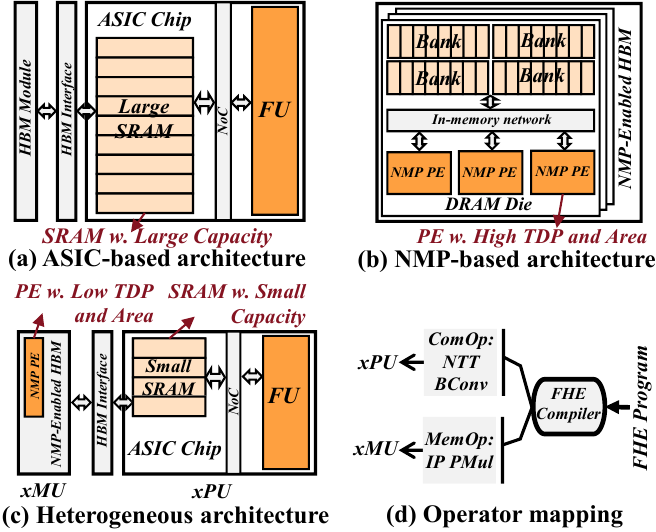}
    \caption{Both ASIC and NMP architectures suffer from high hardware overhead due to large on-chip SRAM (a) and near-memory compute integration (b); the heterogeneous design exploits the strengths of both computation and memory access to achieve a more practical trade-off (c), by mapping ComOps to the ASIC side and MemOps to the NMP side (d). }
    \label{fig:different_architecture}
\end{figure}
Fully Homomorphic Encryption (FHE) has attracted significant attention due to its ability to enable computation on encrypted data in wide privacy-preserving cloud scenarios.
Among existing FHE schemes~\cite{originalckks,bfv,tfhe}, CKKS stands out for supporting approximate arithmetic and efficient SIMD-style computations, making it particularly suitable for privacy-preserving machine learning~\cite{HELR,HEResNet,LoLa,CKKSCNN,BERT}.

CKKS is characterized by both computation-intensive operators (ComOps) with complex computation patterns like ModUp and ModDown, and memory-intensive operators (MemOps) like Inner Product (IP) and Plaintext Multiplication (PMul). Prior works have explored two monolithic accelerator paradigms. First, the ASIC designs~\cite{F1,craterlake,FAB,ARK,SHARP,Trinity,UFC} employ customized modules to accelerate complex ComOps, but require large on-chip memory to alleviate the memory bottleneck, resulting in substantial area and power costs (Fig.~\ref{fig:different_architecture}(a)).
Second, the Near-Memory Processing (NMP) designs~\cite{fhendi,APACHE,FHEmem} leverage high near-memory bandwidth for MemOps but require integrating large computing cores for ComOps (Fig.~\ref{fig:different_architecture}(b)). The integration of a massive complex logic circuit incurs substantial area and power overheads, which hinder its feasibility for implementation~\cite{newton,1ynm}. Both design methodologies face the dilemma of efficiency and hardware costs.

The heterogeneous ASIC-NMP architecture (Fig.~\ref{fig:different_architecture}(c)) is a promising solution to exploit the complementary strengths of both approaches: high-performance ASIC-based modules (xPU) accelerate ComOps, and high-bandwidth NMP modules (xMU) alleviate memory bottlenecks for MemOps.

However, \textit{\textbf{a key challenge of the heterogeneous architecture is the significant communication overhead between the xPU and xMU}} due to the computation pattern of the core operation keyswitch (occupies 80\% computation in CKKS). 
The dataflow of keyswitch consists of an alternating sequence of ComOps and MemOps (ModUp-IP-ModDown), leading to \ding{182} \textit{\textbf{frequent communication when ComOps are mapped onto the xPU and MemOps onto the xMU.}} In addition, the data exchanged between ComOps and MemOps--ciphertexts output from ModUp and IP--are \ding{183} \textit{\textbf{large in size, resulting in considerable single-transfer latency.}} 
Unlike monolithic ASIC designs, where off-chip data can be preloaded to hide latency, the communication for the on-the-fly intermediate results in the heterogeneous architecture lies on the critical path, posing a fundamental bottleneck.
In our evaluation, a heterogeneous accelerator, combining a state-of-the-art monolithic xPU~\cite{SHARP} with bank-level PE-integrated xMU via a 1 TB/s HBM interface, achieves 1.11$\times$ reduction in the bootstrapping computation latency but incurs 8.65$\times$ increase in communication stalls.

In this paper, we propose a communication-light heterogeneous CKKS accelerator with dataflow graph (DFG) optimization and architecture co-design. To address the frequent heterogeneous communication, we propose \textbf{HERO}, a DFG optimization framework to \textit{\textbf{maximize the communication-reduction benefits of hoisting.}} The hoisting algorithm~\cite{doublehoisting} reduces ComOps and communication in heterogeneous systems.
Nonetheless, its benefit is constrained by the limited keyswitch parallelism in the CKKS DFG. 
To circumvent this bottleneck, HERO identifies parallel keyswitch blocks, expands and fuses them to enhance parallelism and maximize hoisting's communication-reduction potential.

To alleviate the on-critical-path communication stall, we propose \textbf{HE\textsuperscript{2}}, an \textbf{\underline{he}}terogeneous F\textbf{\underline{HE}} architecture, \textit{\textbf{utilizing the group-granularity pipeline design to hide the communication latency and improve computation efficiency.}} 
For the xPU, we design dual-level pipelined computation modules to enable both computation-communication overlap and inter-operator overlap for efficient communication hiding.
For the xMU, we design highly parallel yet lightweight modules that exploit the high bank-level bandwidth of HBM.

Our main contributions are as follows:
\begin{itemize}
    \item We construct the first xPU (ASIC)-xMU (NMP) heterogeneous architecture to fully exploit the efficiencies of the xPU for complex ComOps and the high-bandwidth xMU for MemOps, and identify that the bottleneck lies in the intermediate results communication between ComOps and MemOps with detailed profiling.
    \item To address the heavy communication, we propose a co-design that includes a DFG optimization framework for communication-reduction and an inter-group dual-level pipelined hardware for communication-hiding.
    \item Compared to the state-of-the-art CKKS accelerator SHARP~\cite{SHARP}, end-to-end evaluation results show that HE\textsuperscript{2} achieves 1.66$\times$ speedup and 9.23$\times$ EDAP improvement. Specifically, DFG optimization enables hoisting to reduce computation workload and communication volume by an average of 1.64$\times$ and 3.27$\times$, and the pipelined hardware reduces communication stalls to 6.67\%.
\end{itemize}

\section{Background}
\subsection{Basic operators}\label{sec:basic operators}
\begin{table}[]
    \fontsize{12}{18}\selectfont
    \centering
    \caption{Arithmetic Intensity (AI) (Ops per byte) of CKKS Operators. Evaluated under the SHARP~\cite{SHARP}'s parameters. }
    \label{tab:arithmetic intensity}
    \resizebox{1\linewidth}{!}{
    \begin{tabular}{l|cccc|cccc}
        \toprule
         & \multicolumn{4}{c|}{\textbf{Compute Intensive Ops}} & \multicolumn{4}{c}{\textbf{Memory Intensive Ops}} \\ \hline
        \textbf{Op} & \textbf{(I)NTT} & \textbf{BConv} & \textbf{ModUp} & \textbf{ModDown} & \textbf{IP} & \textbf{PMul} & \textbf{CAdd} & \textbf{Rescale} \\ \hline
        \textbf{AI} & 0.89 & 1.60 & 3.38 & 2.92 & 0.12 & 0.09 & 0.07 & 0.11 \\
        \bottomrule
    \end{tabular}}
\end{table}
CKKS is composed of fundamental polynomial-level operators, which are the primary module for acceleration, including computation-intensive (ComOps) and memory-intensive operators (MemOps), with arithmetic intensity shown in Table~\ref{tab:arithmetic intensity}. 

\textbf{NTT} (Number-Theoretic Transform) transforms an input polynomial of degree $N$ from the coefficient to the slot domain through $O(N\ log\ N)$ butterfly operations. 

\textbf{BConv} (Basis Conversion) transforms polynomials within one RNS basis into another. By performing a constant multiplication under the original basis with $l_1$ moduli, followed by another constant multiplication and reduction under $l_2$ target moduli bases, BConv incurs a complexity of $O(l_1\cdot l_2\cdot N)$. 

\textbf{IP} (Inner Product) involves multiplying a group of polynomials with two groups of \textit{evk} polynomials, followed by a sum reduction. IP incurs a substantial memory footprint and a small computation intensity.

\textbf{EWO} (Element-Wise Operation) encompasses operations such as ciphertext-ciphertext addition (CAdd), plaintext-ciphertext addition (PAdd), and multiplication (PMul). EWO exhibits low computation intensity and only involves element-wise addition or multiplication.

\textbf{Autom} (Automorphism) permutes the polynomial, with each coefficient index $i$ mapped to $ik \bmod N$. 

\subsection{Critical primitive in CKKS}
\subsubsection{keyswitch}
In keyswitch~\cite{bv_ghs_hybrid}, the ciphertext under modulus $Q$ is decomposed into \textit{dnum} groups, lifted to $PQ\cdot \textit{dnum}$ via \textbf{ModUp}, multiplied with the \textit{evk} (\textbf{IP}), and reduced back to $Q$ by \textbf{ModDown}. This result is added to the original ciphertext to complete the keyswitch. In CKKS, both ciphertext multiplication and rotation depend on this core primitive.  

\subsubsection{Modulus-Commutative Property} \label{sec:preserving property}
All EWOs and Autom can exchange their execution order with ModUp and ModDown~\cite{doublehoisting,MAD,Anaheim}. We denote such operators as \textbf{\textit{Commutative Operators}}. Taking PMul and CAdd as examples, for ciphertexts $ct$, $ct'$, and plaintext $pt$, the following Equations~\eqref{eq:linear property1} and~\eqref{eq:linear property2} hold. Note that in the right-hand side of Equation~\eqref{eq:linear property1}, the plaintext must be lifted via PModUp as described in~\cite{MAD} to stay in the same domain as ModUp($ct$). 
\begin{equation}\label{eq:linear property1}
    \text{ModUp}(\text{PMul}(ct, pt)) = \text{PMul}(\text{ModUp}(ct),\text{PModUp}(pt))
\end{equation}
\begin{equation}\label{eq:linear property2}
    \text{ModUp}(\text{CAdd}(ct, ct')) = \text{CAdd}(\text{ModUp}(ct), \text{ModUp}(ct'))
\end{equation}

\subsection{Algorithmic Optimizations for PKB}\label{sec:algorithmic optimizations for PKB}
\begin{figure}[]
    \centering
    \includegraphics[width=0.9\linewidth]{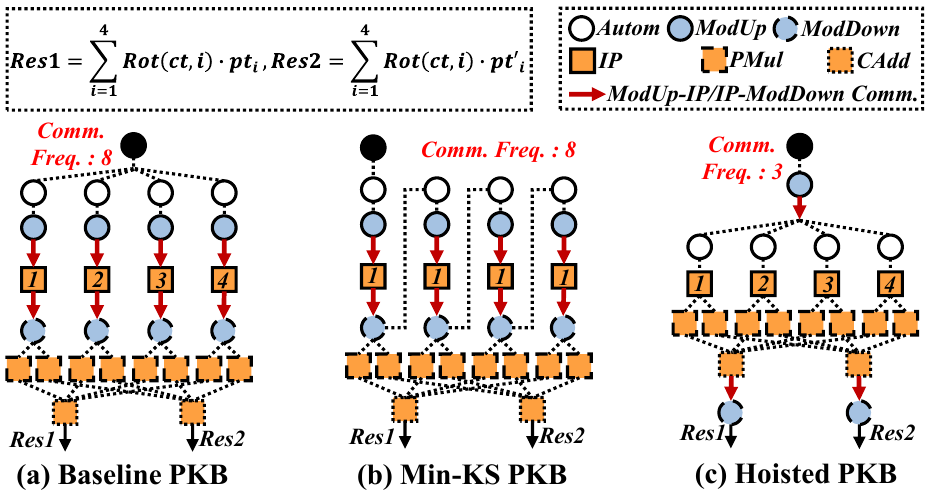}
    \caption{The original PKB performs eight ComOp-MemOp communications (a). Min-KS reduces the number of \textit{evk} but not the communication frequency (b). Hoisting removes redundant ModUps and ModDowns, lowering the ModUp$\rightarrow$IP and IP$\rightarrow$ModDown communications, respectively (c). }
    \label{fig:min-ks double hoisting}
\end{figure}
Multiplication between a plaintext matrix and a ciphertext vector is the key computation in machine learning and C2S/S2C of bootstrapping~\cite{originalckks}. In state-of-the-art implementations~\cite{betterboot,improvedboot}, it uses parallel keyswitches with varying rotation steps, followed by linear combinations via PMul and CAdd. We define blocks of parallel keyswitches as PKBs, as shown in Fig.~\ref{fig:min-ks double hoisting}(a). In plaintext-matrix-ciphertext multiplication, the rotation steps in PKB form an arithmetic progression. 

\textbf{Baby-Step Giant-Step (BSGS)}~\cite{betterboot,ARK} algorithm divides the PKB into groups of smaller PKBs shown in Equation~\eqref{eq:bsgs} and reduces both the computational complexity and the number of \textit{evk} from $O(n_1\cdot n_2)$ to $O(n_1+n_2)$. 
\begin{equation}
\begin{aligned}\label{eq:bsgs}
    &\qquad\,\,\,\,\,\,\,\sum_{i=1}^{n_1\cdot n_2} \text{PMul}(\text{Rot}(ct,s_i), pt_i) \\
   = &\overbrace{\sum_{{i}=1}^{n_1}\text{Rot}(\underbrace{\sum_{j=1}^{n_2}\text{PMul}(\text{Rot}(ct,s_j), pt_{i,j})}_{\text{Baby step}}, s_i)}^{\text{Giant step}}
\end{aligned}
\end{equation}

The \textit{\textbf{Min-KS}}~\cite{ARK} approach further reduces the required \textit{evk} number by converting the PKB to serial rotations to ensure that each rotation step is uniform (Fig.~\ref{fig:min-ks double hoisting}(b)). 

The \textit{\textbf{hoisting}} technique~\cite{doublehoisting} exchanges ModUps/ModDowns with commutative operators and merges those from different keyswitches, as shown in Fig.~\ref{fig:min-ks double hoisting}(c). By reducing the number of ComOps, hoisting lowers communication frequency between ComOps and MemOps. However, while it lessens the ComOps workload, it increases that of MemOps, whose computation order is swapped, shifting their modulus domain from $Q$ to $PQ$ or $PQ\cdot \textit{dnum}$. 

\section{Motivation}
\subsection{Existing ASIC-/NMP-based Architectures Face Dilemma of Performance and Hardware Costs}
\begin{figure}[]
    \centering
    \includegraphics[width=1\linewidth]{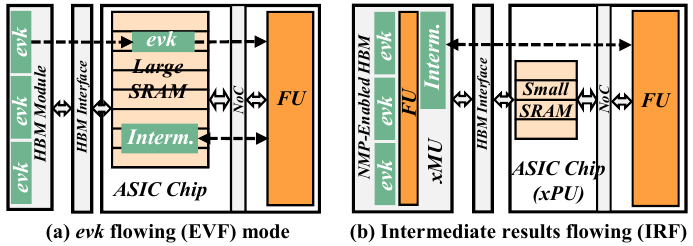}
    \caption{Two dataflow schemes exist in an ASIC chip equipped with HBM. An \textit{evk} flowing dataflow preloads \textit{evk} to the chip for IP computation (a). An intermediate results flowing scheme transfers intermediates of keyswitch to xMU for IP computation, where data transfers are on the critical path (b).}
    \label{fig:xPU-xMU dataflow}
\end{figure}
The coexistence of compute-intensive operators (ComOps) and memory-intensive operators (MemOps) in CKKS poses major challenges for hardware-efficient acceleration. Prior studies have explored ASIC-based~\cite{F1,FAB,BTS,craterlake,ARK,SHARP,Trinity,alchemist} and NMP-based~\cite{fhendi,APACHE,FHEmem} monolithic architectures, both incurring substantial hardware overhead. 
\textbf{First}, ASIC accelerators achieve efficient ComOps acceleration with low area and power consumption by using advanced process nodes~\cite{7nm} and employ customized computation modules. However, CKKS workloads demand large on-chip SRAM for MemOps, which can consume up to half of the total area and power consumption in ASIC accelerators~\cite{ARK,BTS}. 
\textbf{Second}, NMP designs exploit high near-memory bandwidth to accelerate MemOps and integrate deeply pipelined compute units within low hierarchies in DRAM for ComOps~\cite{fhendi,flexmem}, but this results in complex and costly designs. 
Specifically, pure in-die NMP cannot provide adequate computational capability for FHE within the area budget (25\% suggested by SK Hynix AiM~\cite{newton,1ynm}), owing to the scarcity of logic resources in DRAM's technology that transistors operate roughly 3$\times$ slower and the density is 10$\times$ lower than CMOS at the same technology node~\cite{dram_scarce_compute}. 
Additionally, the support for complex FHE ComOps leads to considerable power overhead, necessitating customized high-end-server active heat sink thermal management~\cite{fhendi,hbmpower}, thereby further complicating industrial tape-out.
These two paradigms inspire us to propose a hardware-efficient ASIC (xPU)-NMP (xMU) heterogeneous architecture featuring an xPU with small on-chip memory and an xMU supporting only simple MemOps, which introduces low extra logic integration. 

\subsection{Frequent and Large Communication Traffic is the Key Challenge in Heterogeneous Architectures}\label{sec:communication challenges}
According to prior ASIC designs~\cite{SHARP,BTS}, large on-chip SRAM stores preloaded \textit{evks} for IP and baby-step ciphertexts for PMul in the BSGS phase of bootstrapping. To lower the on‑xPU memory demand, an efficient mapping of IPs and PMuls is essential.

IP appears exclusively in the keyswitch. In an xPU-xMU heterogeneous system, keyswitch admits two dataflow design choices: the \textit{evk flowing} (EVF) dataflow used in existing ASIC monolithic accelerators (Fig.~\ref{fig:xPU-xMU dataflow}(a)) and a new \textit{intermediate results flowing} (IRF) dataflow (Fig.~\ref{fig:xPU-xMU dataflow} (b)). The EVF dataflow offloads the entire keyswitch to the xPU. It allows \textit{evk} to be reused and preloaded to reduce the cost of off-chip memory access. However, when the program exhibits sequential keyswitches with low \textit{evk} reuse, its performance remains constrained by large on-xPU memory~\cite{ARK}. 
The IRF dataflow maps IPs to the xMU to avoid loading \textit{evk} to the xPU. In this mode, the ModUp output is sent to the xMU for IP, and the result is then returned to the xPU for ModDown.

\begin{figure}[]
    \centering
    \includegraphics[width=1\linewidth]{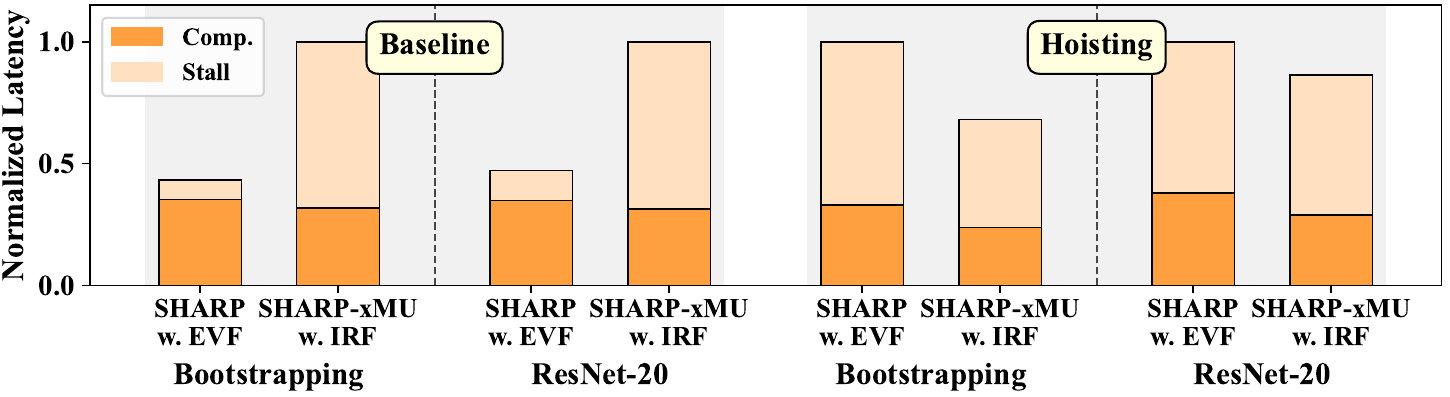}
    \caption{Baseline adopts Min-KS as in~\cite{ARK,SHARP}. SHARP-xMU is an xPU-xMU heterogeneous accelerator, in which the xPU is aligned with SHARP with detailed configurations provided in Sec.~\ref{sec:he2 evaluation}. Communication is the primary bottleneck in IRF‑based heterogeneous architectures, and hoisting can partially reduce communication overhead.}
    \label{fig:comm impact}
\end{figure}

Although IRF reduces xPU memory use and exploits xMU near-memory bandwidth, it introduces two additional data transfers per keyswitch, i.e., between ModUp IP and IP ModDown. Considering the dominance of keyswitch in CKKS, \ding{182} \textbf{\textit{the communication between xPU and xMU becomes frequent and degrades the overall performance when IRF is applied.}} 
Furthermore, this communication lies on the critical path, involving up to 144 MB of intermediate ciphertext per transfer. We extend a state‑of‑the‑art ASIC design~\cite{SHARP} by integrating an xMU component into a heterogeneous xPU‑xMU architecture. In this architecture, \ding{183} \textbf{\textit{the existing xPU design cannot effectively balance communication and computation, making it difficult to hide the latency of intermediate result transfers along the critical path.}} In our experiments, this communication stall accounts for 68.2\% and 68.7\% of the total latency in bootstrapping~\cite{improvedboot} and ResNet-20~\cite{HEResNet}, as shown in the left region of Fig.~\ref{fig:comm impact}.

\subsection{Communication Optimization Potential of Hoisting and Its Limitation under Low Keyswitch Parallelism}\label{sec:limited hoisting scope}
Hoisting~\cite{doublehoisting} extracts shared ModUps and ModDowns from multiple parallel keyswitches. Although the corresponding \textit{evks} differ, the ModUp results can be reused across IPs after hoisting, and the aggregated results from multiple IPs require only one ModDown. Thus, hoisting trades \textit{evk} reuse for improved reuse of intermediate results.

\begin{figure}[]
    \centering
    \includegraphics[width=1\linewidth]{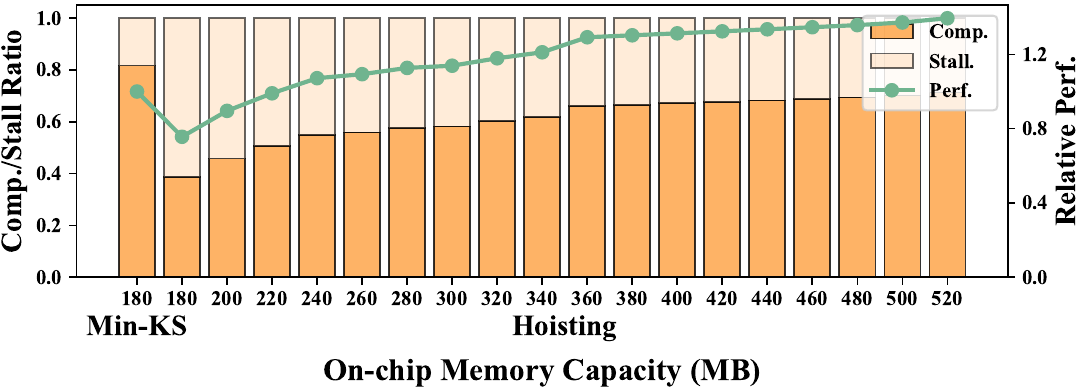}
    \caption{Bootstrapping performance of SHARP with hoisting across varying on-chip memory capacities. Increasing on-chip memory allows \textit{evk} preloading to mitigate off-chip stalls, but reduced \textit{evk} reuse under hoisting yields diminishing performance gains.}
    \label{fig:sharp with hoist}
\end{figure}

\textbf{\textit{Due to reduced \textit{evk} reusability, hoisting causes memory access stalls and performance loss in EVF‑based architectures.}} As depicted in Fig.~\ref{fig:sharp with hoist}, directly applying hoisting to SHARP leads to performance degradation and offers only a 39.4\% speedup despite a 2.89$\times$ increase in on-chip memory. 
In contrast, \textbf{\textit{hoisting improves intermediate result reuse, reducing communication frequency in IRF-based heterogeneous architectures and thus boosting performance.}} Consequently, the IRF‑based system achieves moderately higher performance than its EVF‑based counterpart when hoisting is applied, as shown in the right region of Fig.~\ref{fig:comm impact}. Moreover, hoisting enables PMul in BSGS of the bootstrapping to be reordered and offloaded to the xMU along with adjacent IPs (Fig.~\ref{fig:min-ks double hoisting}). Consequently, the substantial on-chip memory footprint of baby-step ciphertexts described in Sec.~\ref{sec:communication challenges} can be avoided. For bootstrapping and ResNet-20, hoisting can reduce the communication stall of IRF-based systems by 2.00$\times$ and 1.61$\times$ compared to the two baseline benchmarks, respectively. 

\begin{figure}[]
    \centering
    \includegraphics[width=1\linewidth]{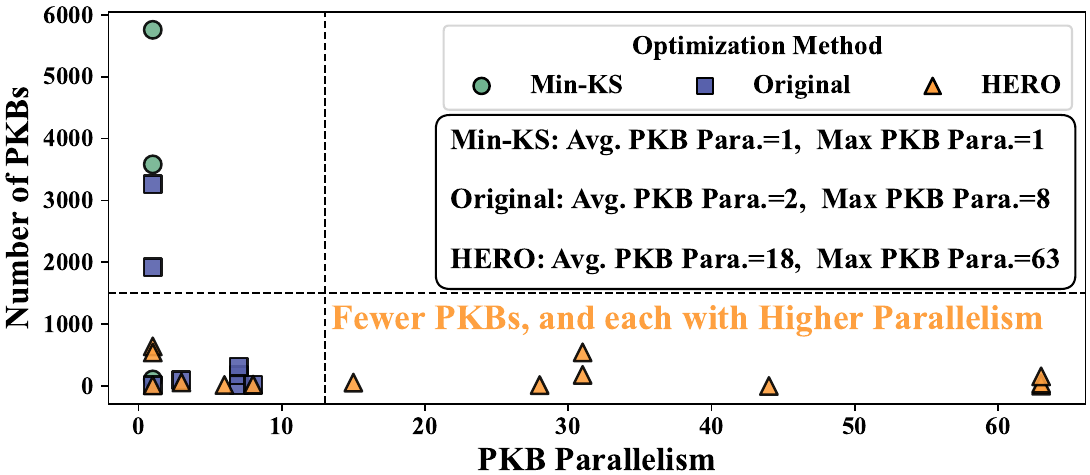}
    \caption{Parallelism and number of PKBs in the bootstrapping, ResNet20, and HELR. CKKS program contains numerous segments with low keyswitch parallelism. Min‑KS further increases the number of PKBs while producing PKBs with lower parallelism. By applying our algorithmic optimization introduced in Sec.~\ref{sec:HERO}, low-parallelism PKBs are fused into a smaller number of highly parallel PKBs, thereby unveiling the benefits of hoisting.}
    \label{fig:parallelism}
\end{figure}

Nevertheless, \ding{184} \textbf{\textit{the optimization effect of hoisting on communication is limited due to the low keyswitch parallelism of CKKS programs.}} Specifically, through hoisting, redundant ModUps and ModDowns within the PKB can be reduced to a level proportional to the PKB’s input and output degrees. Since the saved communication volume scales with the reduction in ModUp and ModDown, a PKB with higher parallel keyswitches and lower in/out‑degree can achieve more communication savings. In CKKS programs, many fragmented PKBs with a limited keyswitch-parallelism (less than 10) exist, as shown in Fig.~\ref{fig:parallelism}, which restricts the hoisting effect. 
Therefore, in Sec.~\ref{sec:HERO}, we propose a program graph level optimization method (HERO) to fuse multiple low-parallelism PKBs into a few PKBs with higher parallelism (more than 30), to fully exploit the potential of hoisting. 

\subsection{New Heterogeneous-Specific Performance Model is Needed}
\begin{figure}[]
    \centering
    \includegraphics[width=1\linewidth]{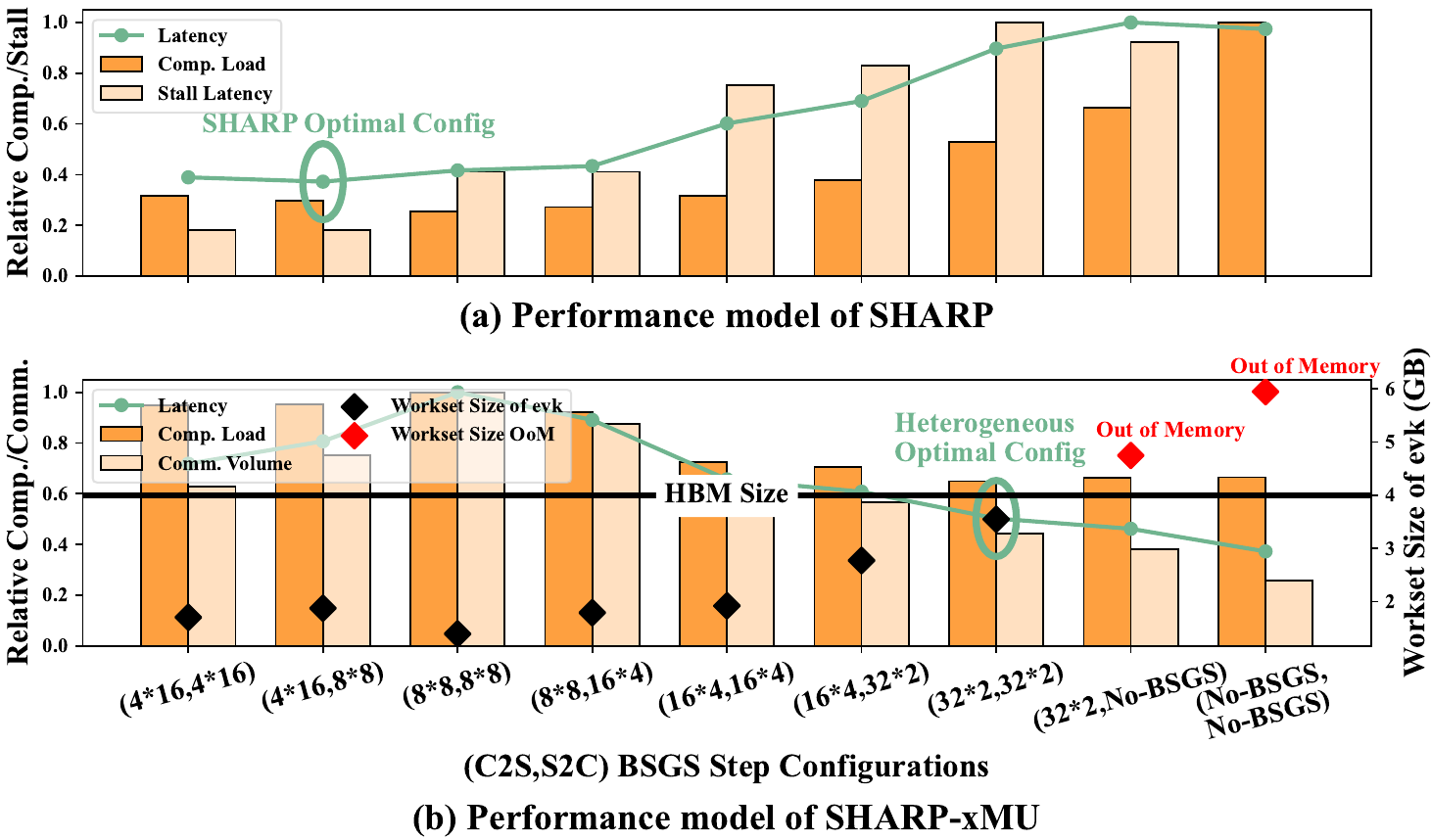}
    \caption{BSGS parameter exploration. EVF-based monolithic architectures (e.g., SHARP) consider computation load and off-chip memory access stalls. In contrast, the performance model of IRF-based heterogeneous accelerators (e.g., SHARP-xMU) incorporates computation load, communication volume, and the \textit{evk} workset size. The difference in the considered factors leads to distinct optimal parameter configurations.}
    \label{fig:performance model}
\end{figure}

Since this work marks the first attempt at an ASIC‑NMP heterogeneous acceleration for CKKS, and prior studies have not thoroughly explored the potential of hoisting, we observe that the performance models for existing monolithic CKKS accelerators, which are primarily based on computation volume and off-chip memory access stall under the EVF dataflow, are insufficient to accurately assess the impact of hoisting and IRF in heterogeneous systems. 
In prior work on BSGS of bootstrapping with EVF dataflow, the baby-step (\textit{bs}) and giant-step (\textit{gs}) are jointly optimized to minimize computation cost and off‑chip memory traffic. The computation cost is lowest when \textit{bs} and \textit{gs} are equal (i.e., both 8), but this exceeds the on‑chip memory for baby-step ciphertexts~\cite{SHARP}, causing frequent intermediate swaps. Thus, the optimal configuration of~\cite{SHARP} adopts a smaller \textit{bs} of 4 (Fig.~\ref{fig:performance model}(a)). 

In contrast, with hoisting and IRF, choosing larger gaps between \textit{bs} and \textit{gs} exposes more parallelism for hoisting, reducing communication and computation. However, such configurations increase the \textit{evk} storage demand and may exceed HBM capacity, as depicted in Fig.~\ref{fig:performance model}(b). 
This motivates the need to \ding{185} \textbf{\textit{develop a new heterogeneous‑specific performance model that jointly considers computation, communication, and the size of \textit{evk} working set for general CKKS applications.}}

\section{HERO: A Hoisting-Enhanced DFG Optimization Framework}\label{sec:HERO}
\begin{figure*}[]
    \centering
    \includegraphics[width=1\linewidth]{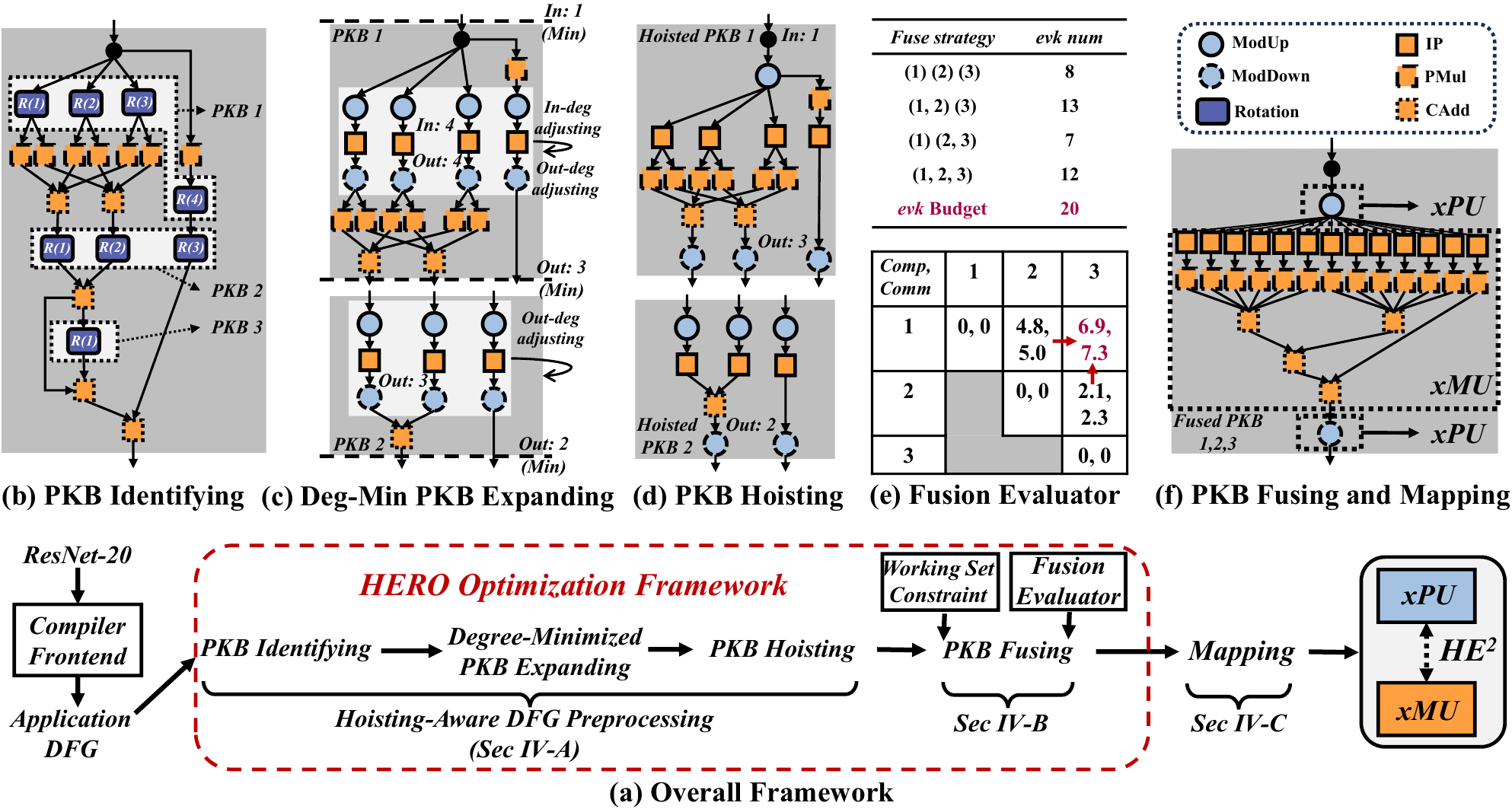}
    \caption{The overall procedure of HERO (a). The PKB in CKKS DFG is first identified (b), followed by expanding each PKB for minimizing the degree (c) to enhance ComOp-MemOp communication reduction by hoisting and placing redundant ModUps and ModDowns to positions with minimized degrees (d). The fusion evaluator assesses the computation and communication reduction from fusing hoisted PKBs according to the relative MMul numbers and intermediates' size, along with the required \textit{evk} counts (e), ultimately selecting the optimal fusing strategy under the storage capacity constraints (f). Identical indices in different rotation nodes do not imply the same rotation step size. }
    \label{fig:framework}
\end{figure*}

To tackle the critical communication challenges on heterogeneous systems, and to explore the optimization potential of hoisting, we abstract dataflow graphs (DFGs) from applications, where nodes represent operators and edges denote data dependencies. Based on this, we propose \textbf{HERO}, a \underline{\textbf{h}}oisting-\underline{\textbf{e}}nhanced communication \underline{\textbf{r}}eduction DFG \underline{\textbf{o}}ptimization framework. The input to HERO is a DFG generated by an FHE compiler~\cite{resbm,chet,eva}. To enable the generality of hoisting in arbitrary CKKS programs, we identify and locally expand PKBs for degrees fine-tuning (Sec.~\ref{sec:preprocessing}). This preprocessing stage reshapes the local DFG, thereby enhancing the effectiveness of hoisting on subgraphs. Furthermore, to enhance the keyswitch-level parallelism that hoisting can leverage, we introduce, for the first time, a novel technique called PKB fusing, which fuses PKBs at the global DFG level and leverages a fusion evaluator to derive the optimal fusion scheme (Sec.~\ref{sec:Fusing PKBs}). Finally, we map the restructured DFG onto the proposed heterogeneous accelerator based on the parallelism of each PKB (Sec.~\ref{sec:dataflow mapping}). 

\subsection{Preprocessing: Subgraph Degrees Minimizing for Locally Optimal Hoisting}\label{sec:preprocessing}

Existing hoisting-based studies~\cite{Anaheim,FAST,MAD} achieve better hoisting effects by adjusting commutative operators to reduce the degrees of PKBs during the bootstrapping. Our preprocessing procedure performs similarly, but it is not limited to a certain application~\cite{orion}. Instead, we propose a general method by formalizing this process into the following two steps, making it applicable to PKBs within arbitrary CKKS programs, as illustrated in Fig.~\ref{fig:framework}(a) and (b).

\noindent\textit{\textbf{PKB identifying.}} We traverse the DFG from the inputs, assign keyswitches to layers based on their order along each path, and group those in the same layer into PKBs, yielding a partitioned DFG composed of sequential PKB layers and intermediate operations, as shown in Fig.~\ref{fig:framework}(a).

\noindent\textit{\textbf{Degree-minimized PKB expanding.}}
Hoisting places the ModUp and ModDown of a PKB at its input and output, respectively. To maximize its benefit (i.e., reduce the number of ModUps and ModDowns), we minimize each PKB's in- and out-degree by greedily expanding it with commutative operators, and apply hoisting as shown in Fig.~\ref{fig:framework}(b) and (c). 

As noted in Sec.~\ref{sec:preserving property}, this expansion enlarges the modulus domain and thus increases the cost of MemOps (e.g., PMul, CAdd), but the overhead is generally outweighed by the reduced ModUps and ModDowns. 

\subsection{PKB Fusing for Hoisting Efficiency Enhancement}\label{sec:Fusing PKBs}
Although each PKB is locally optimized through subgraph degrees minimizing, the parallelism of keyswitch within CKKS programs remains unchanged and low (Fig.~\ref{fig:parallelism}), which limits the overall gains of hoisting. 

Unlike prior works~\cite{FAST,MAD,Anaheim} that apply hoisting without modifying the underlying CKKS program, we further enhance its benefits by proposing \textit{PKB Fusion}, which enlarges the parallelism and reduces the in- and out-degree of PKBs. Specifically, consider two adjacent PKBs with $n_1$ and $n_2$ rotations, separated by EWOs $\text{F}_i$, where each rotation in the second PKB depends on all rotations in the first. We perform a transformation analogous to the inverse BSGS process (Equation~\eqref{eq:PRB fusion}). Since EWOs such as PMul can commute with rotations, i.e., $\text{Rot}(\text{PMul}(ct,pt))=\text{PMul}(\text{Rot}(ct),\text{Autom}(pt))$, we traverse each of the $n_2$ paths in the second PKB backward and push EWOs after the rotations. This makes each of the $n_2$ paths directly adjacent to one of the $n_1$ paths in the first PKB. We then exploit the additive property of rotations, i.e., $\text{Rot}(\text{Rot}(ct,s),t) = \text{Rot}(ct,s+t)$, to merge the two consecutive rotations into a single one. After pairing every rotation path in the second PKB with all $n_1$ paths in the first PKB, the two serial PKBs are fused into a larger PKB with $O(n_1\cdot n_2)$ rotations, where the output nodes of PKB1 and input nodes of PKB2 are removed. Consequently, hoisting on this fused PKB eliminates additional ComOps and yields more savings in both computation and communication. 

\begin{equation}
\begin{aligned}\label{eq:PRB fusion}
    &\overbrace{\{\text{Rot}(\underbrace{\text{F}_i(\{\text{Rot}(ct,s_j)\}_{1 \leq j \leq n_2})}_{\text{The first PKB}}, s_i')\}_{1\leq i\leq n_1}}^{\text{The second PKB}} \\= 
    &\{\text{F}_i'(\{\text{Rot}(\text{Rot}(ct,s_j),s'_i)\}_{1 \leq j \leq n_2})\}_{1\leq i\leq n_1} \\=
    &\underbrace{\{\text{F}_i'(\{\text{Rot}(ct,s_j+s'_i)\}_{1 \leq j \leq n_2})\}_{1\leq i\leq n_1}}_{\text{The Enlarged PKB}}
\end{aligned}
\end{equation}

However, this comes with a trade-off: while hoisting becomes more beneficial, the number of \textit{evks} increases, the amount of IPs grows, and additional MemOp overheads are introduced. 
Therefore, \textit{\textbf{when applying PKB fusing, it is necessary to balance computation cost, communication overhead, and the working set size of evks.}}

\noindent\textit{\textbf{Formalization.}} The impact of PKB fusion can be modeled as follows. We consider two PKBs to be fused: PKB1 with out-degree $outdeg_1$ and $n_1$ rotations, and PKB2 with in-degree $indeg_2$ and $n_2$ rotations. After fusion, on the one hand, hoisting eliminates extra $outdeg_1$ ModDowns and $indeg_2$ ModUps, along with the associated communication volume. On the other hand, fusion introduces additional (1) up to $n_1\cdot n_2 - n_1 - n_2$ IPs and \textit{evk} storage overhead, with the actual number determined by the non-duplicated subset among $n_1\cdot n_2$ keys, (2) the extra computation incurred by increasing the modulus of the EWOs between two PKBs. 

\begin{figure}[]
    \centering
    \includegraphics[width=1\linewidth]{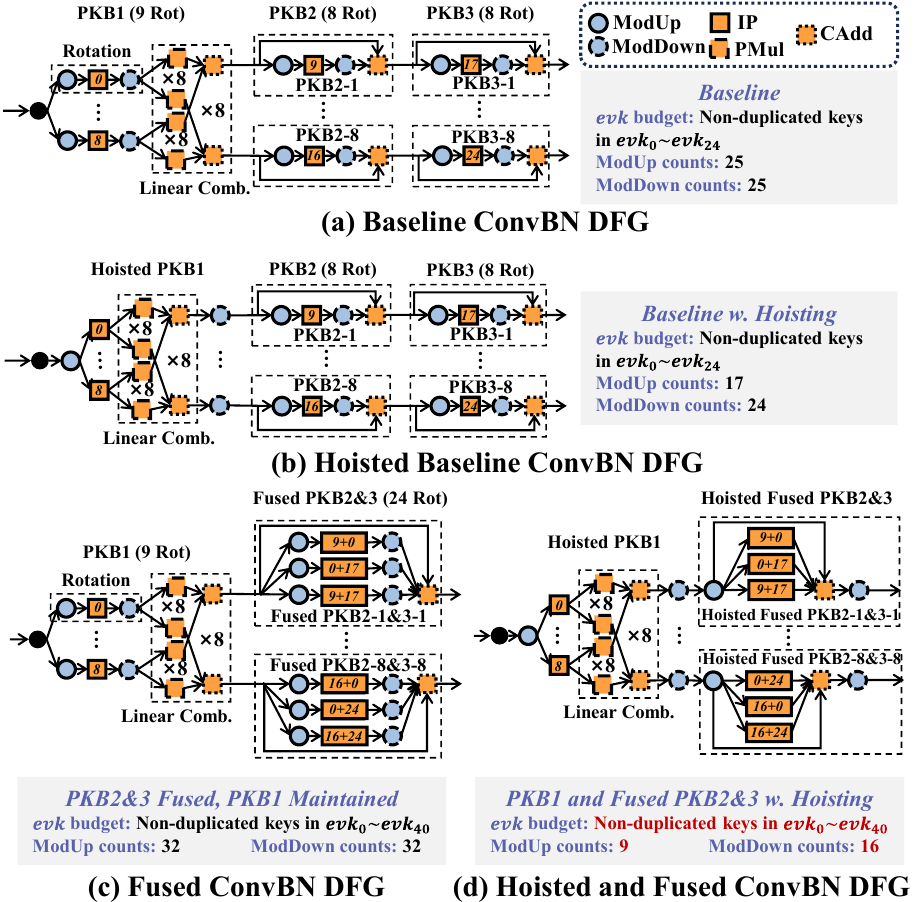}
    \caption{PKB fusion example. The index inside the IP node represents a schematic value of the rotation step, rather than the actual step in real cases. As the number of fused PKBs increases, the fused PKB contains more parallel rotations, which requires a larger number of \textit{evks} and shows more ComOps reduction with hoisting. }
    \label{fig:fusing option}
\end{figure}

\noindent\textit{\textbf{Case Study.}} We illustrate the working mechanism and effects of PKB fusion through a simplified example of the ConvBN DFG in~\cite{HEResNet}. Fig.~\ref{fig:fusing option}(a) shows a DFG containing three PKBs, each with 9, 8, and 8 parallel rotations, respectively. In this configuration, the number of \textit{evks} corresponds to the non-duplicated subset of 25 keys, with both the numbers of ModUps and ModDowns being 25. Fig.~\ref{fig:fusing option}(b) demonstrates the effect of applying hoisting to the raw DFG, as done in Anaheim~\cite{Anaheim} and FAST~\cite{FAST}. In this case, the parallel ModUps in PKB1 are reduced from 9 to 1, and the ModDowns can be extracted after linear combination and reduced from 9 to 8. However, for PKB2 and PKB3, hoisting does not provide further optimization, as the parallel rotations in these PKBs lack common predecessor or successor nodes, thereby preventing any reduction in ModUps/ModDowns via hoisting. 
In this case, because hoisting has no impact on the number of IPs, the number of \textit{evks} is changed.

In Fig.~\ref{fig:fusing option}(c), PKB2 and PKB3 are fused into an enlarged PKB. During fusion, all data-dependent paths are combined according to the additive property of rotation steps, as formulated in Equation~\ref{eq:PRB fusion}, where $n_1$ and $n_2$ represent the numbers of data-dependent paths. 
Specifically, PKB2‑1 and PKB3‑1 contain dependent paths. In PKB2‑1, two parallel paths are summed to produce the input to PKB3‑1, which is subsequently processed through its own two parallel paths. When PKB2‑1 and PKB3‑1 are fused, their respective two parallel paths are combined via rotation‑step addition, resulting in a fused PKB comprising four parallel paths, three of which correspond to parallel rotations.
After fusion, the $n_1+n_2$ rotations described in Equation~\eqref{eq:PRB fusion} become $O(n_1\cdot n_2)$ rotations, leading to an increased PKB parallelism. In this case, by applying hoisting as shown in Fig.~\ref{fig:fusing option}(d), the high-parallel ModUps and ModDowns can be reduced and extracted to the front and end of the eight parallel paths of the fused PKB, without performing additional ModUps and ModDowns between PKB2 and PKB3 as in Fig.~\ref{fig:fusing option}(b).

This example demonstrates that the combination of PKB fusion and hoisting effectively reduces the number of ModUps and ModDowns at the cost of increased \textit{evk} working set and MemOp computation volume.

\noindent\textit{\textbf{Fusion Evaluator.}} To evaluate the benefit of fusing under storage constraints, we define $\textit{FuseScore}(i,j)$ as follows: if the number of \textit{evks} after fusing PKB $i$ and PKB $j$ exceeds the available storage capacity, the score is marked invalid; otherwise, for PKB $i$ with $n_i$ rotations and PKB $j$ with $n_j$ rotations, \textit{FuseScore} denotes the maximal combined computation and communication savings among all consistent PKB pairs with an invariant product $n_i\cdot n_j$.
\begin{equation}
\begin{aligned}\label{eq:DP}
    \textit{DP}[i][j] = &\max_{1\leq j' \leq j-1}\textit{DP}[i][j'] + \textit{DP}[j'+1][j]\\
    & + \textit{FuseScore}(j',j'+1) 
\end{aligned}
\end{equation}
We formulate the global DFG fusion as a dynamic programming (\textit{DP}) problem. A two-dimensional \textit{DP} table is constructed, where diagonal entries represent the base case of unfused PKBs and are initialized to zero. Each entry $(i,j)$ stores the optimal cumulative \textit{FuseScore} for covering PKB $i$ through PKB $j$. The table is populated iteratively toward the upper-right corner according to the transition equation~\eqref{eq:DP}, and the resulting solution yields the globally optimal fusion plan that minimizes computation and communication overhead while satisfying storage constraints. Ultimately, compared with applying hoisting in unmodified programs, HERO achieves a 2.25$\times$ more reduction in computation load and 2.42$\times$ more reduction in communication volume. Detailed evaluation is in Sec.~\ref{sec:impact of DFG optimization}.

\subsection{BSGS Configuration Exploration}\label{sec:bsgs in bootstrap}
The Baby-Step Giant-Step (BSGS) algorithm is a key technique to reduce the computational cost in the C2S and S2C phases of bootstrapping. C2S/S2C involves a PKB with \textit{D} parallel rotations, followed by PMuls and CAdds. BSGS restructures this computation into two stages:
\begin{itemize}
    \item \textit{\textbf{Baby step:}} A ciphertext is rotated by \textit{bs} different strides to obtain \textit{bs} results, which are linearly combined via PMul and CAdd to yield $\textit{gs}\coloneqq D/\textit{bs}$ results.
    \item \textit{\textbf{Giant step:}} Each of the \textit{gs} ciphertexts is further rotated once and summed to yield the final result.
\end{itemize}
    
Essentially, BSGS replaces one PKB with parallelism \textit{D} and in/out-degree of 1 into two serial PKBs: \textbf{PKB1} with parallelism \textit{bs}, in-degree 1, and out-degree \textit{bs}; and \textbf{PKB2} with parallelism \textit{gs}, in-degree \textit{gs}, and out-degree 1. Thus, \textit{\textbf{BSGS decreases overall keyswitch parallelism and increases the in/out-degrees of subgraphs.}} When combined with hoisting, PKB1 has one ModUp and \textit{bs} ModDowns, and PKB2 has \textit{gs} ModUps and one ModDown. 
Without BSGS, only one ModUp and one ModDown are required. Although the number of IPs grows from $\textit{bs}+\textit{gs}$ to $O(\textit{bs}\cdot \textit{gs})$, the total overhead tends to decrease. 

In practical applications, HERO selectively disables BSGS when memory suffices for the extra \textit{evk}. When the memory is limited, HERO prefers configurations with a larger gap between \textit{bs} and \textit{gs}, as this can lead to greater reductions in computation and communication overhead with hoisting, as shown in Fig.~\ref{fig:performance model}(b).

\subsection{Dataflow Mapping}\label{sec:dataflow mapping}
For the HERO‑optimized program, its DFG is partitioned into a sequence of PKBs, each starting with ComOps (ModUps), followed by MemOps (IP, PMul, etc.), and ending with ComOps (ModDowns). In addition, EWOs may occur between adjacent PKBs. We consider two alternative mapping schemes. 

\noindent \textbf{\textit{IRF scheme.}} We directly map ComOps and MemOps to the xPU and xMU, respectively, and map inter-PKB EWOs onto the xPU. In this case, all IPs are entirely executed on the xMU, allowing us to minimize on‑xPU storage.

\noindent \textbf{\textit{Hybrid scheme.}} For each PKB, we choose the dataflow based on its IP‑level parallelism: we adopt IRF only when the IP parallelism is greater than 1; otherwise, we adopt EVF dataflow. The hybrid scheme enables us to use EVF when the cost of moving \textit{evks} is lower than that of transferring intermediate results, thereby reducing communication overhead. However, it requires reserving additional on‑xPU storage for one \textit{evk}. We evaluate both schemes in Sec.~\ref{sec:evaluation}.

\section{Heterogeneous Architecture Design}
\subsection{Architecture Overview}
\begin{figure}[]
    \centering
    \includegraphics[width=0.95\linewidth]{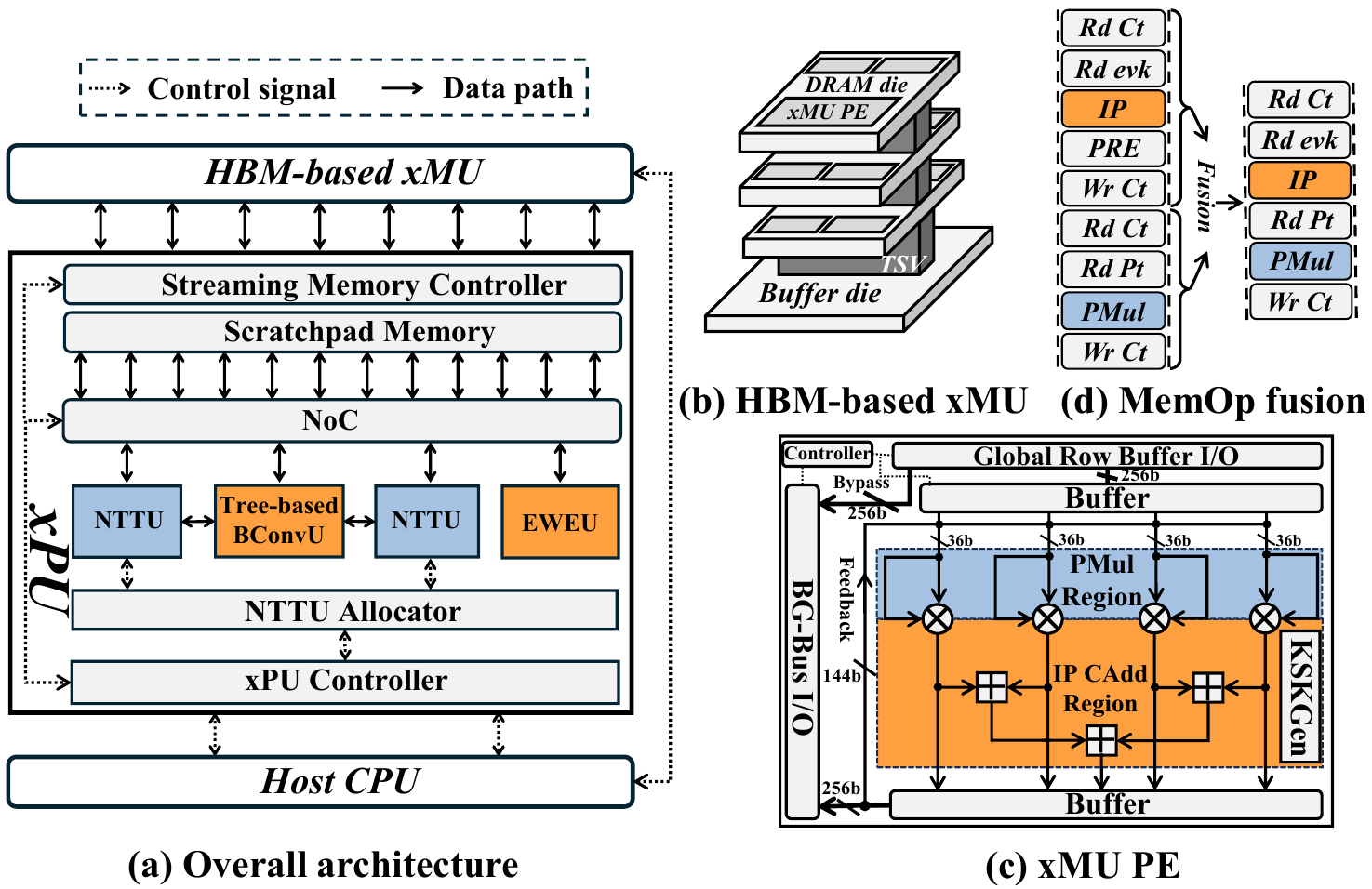}
    \caption{The overall heterogeneous architecture contains an ASIC-based xPU (a) and an HBM-based xMU (b). The xMU PEs are deployed within the column decoder of all HBM banks (c). MemOps performed in xMU are fused to reduce write-back latency (d). }
    \label{fig:overview}
\end{figure}
Although HERO reduces the frequency of heterogeneous communications, the latency of transferring intermediate results during keyswitch remains a bottleneck. The limitation arises from the xPU architecture's design philosophy~\cite{F1,BTS,SHARP,ARK,FAST}, which primarily prioritizes minimizing per-operator latency over balancing overall computation and communication delays. Consequently, the existing microarchitecture fails to simultaneously hide communication latency and sustain high overall performance.

We propose HE\textsuperscript{2}, a low-cost heterogeneous accelerator that delivers high performance and efficiently hides communication stalls. HE\textsuperscript{2} consists of two main core components, as illustrated in Fig.~\ref{fig:overview}. 

\noindent\textit{\textbf{xPU Design.}} We redesign the xPU to reduce per-operator hardware cost and balance operator latency with communication delay. Under limited single‑operator performance, we develop new microarchitectures to enable inter-operator overlap (Sec.~\ref{sec:xPU}). Furthermore, we reconstruct the critical path from INTT$\rightarrow$BConv$\rightarrow$NTT into parallel BConv$\rightarrow$NTT and NTT paths, improving computational parallelism (Sec.~\ref{sec:parallelism exploitation}). 

\noindent\textit{\textbf{xMU Design.}} We propose a low-overhead and efficient xMU design along with the MemOps fusion technique and an in-memory automorphism method (Sec.~\ref{sec:xMU}).

\subsection{Dual-level Pipelining xPU Microarchitecture for Communication-Stall Mitigation}\label{sec:xPU}
\begin{figure}[]
    \centering
    \includegraphics[width=1\linewidth]{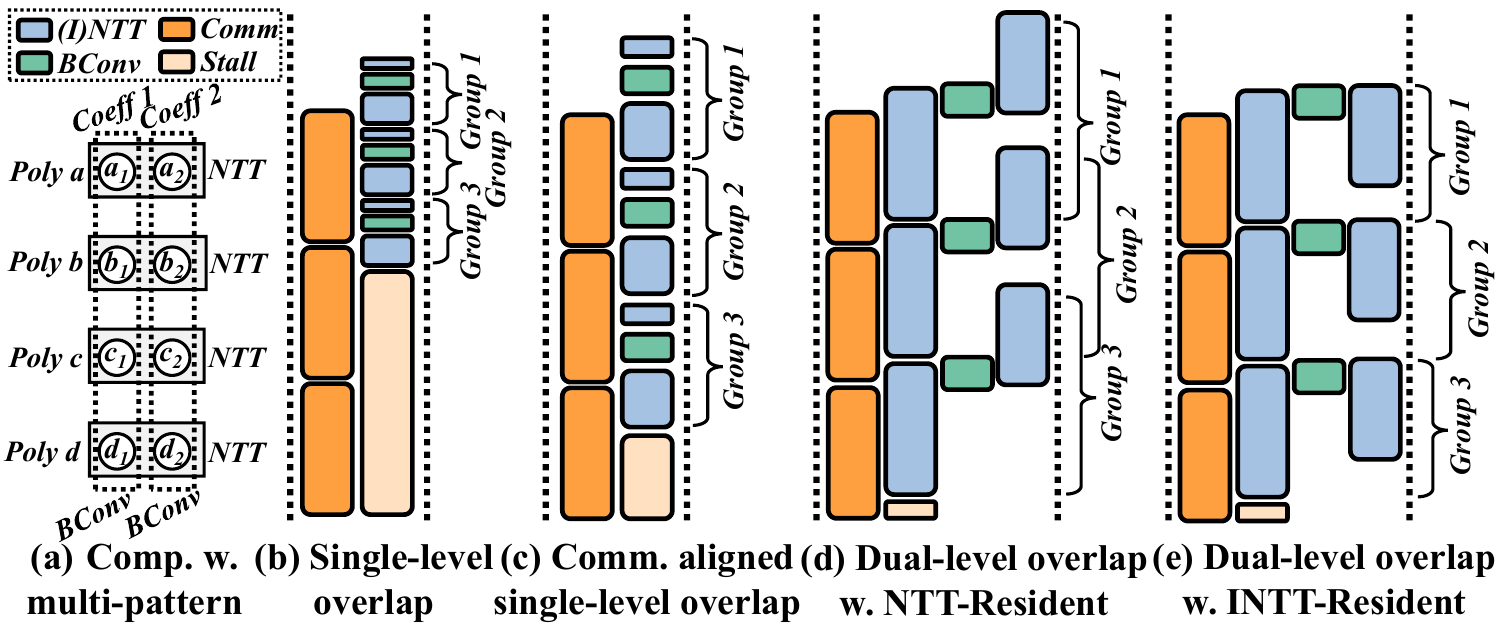}
    \caption{Distinct computation patterns of NTT and BConv prevent efficient overlap (a). In a ModUp pipeline with 3 groups, the previous architecture~\cite{SHARP} shows low overlap for both computation and communication (b). Na\"ively reducing its parallelism partially mitigates communication stalls but at the cost of performance degradation (c). Dual-level overlapped xPU architecture hides substantial communication stalls (d), and the design of INTT-Resident keyswitch pipeline avoids performance degradation (e).}
    \label{fig:dimension}
\end{figure}
Compared with the monolithic architecture adopting a consistent EVF dataflow, HE\textsuperscript{2} switches into an IRF dataflow for hoisted PKBs, where data transfers lie on the critical path of the keyswitch, exposing long communication stalls. 
Since MemOps executed in the near-memory module occupy the HBM ports and block concurrent xPU-HBM transfers and MemOps computation, we overlap communication and the computation on xPU, typically ModUp and ModDown.

ModUp and ModDown process multiple decomposed ciphertext groups, each following an INTT$\rightarrow$BConv$\rightarrow$NTT pipeline. Inter-group parallelism allows na\"ive computation communication overlap, where data transfers for one group are pipelined with computations for others. However, existing ASICs~\cite{BTS,FAB,ARK,SHARP} stack highly parallel and low-latency compute units, thus suffer from communication bottlenecks and underutilized compute resources, as shown in Fig.~\ref{fig:dimension}(b). Simply lowering computation parallelism cannot remove these stalls, as depicted in Fig.~\ref{fig:dimension}(c). To overcome this, we design an xPU with \textit{\textbf{dual-level overlapping}}, combining computation-communication and inter-operator overlap to sustain acceleration performance while hiding communication latency. 

\begin{figure}[]
    \centering
    \includegraphics[width=1\linewidth]{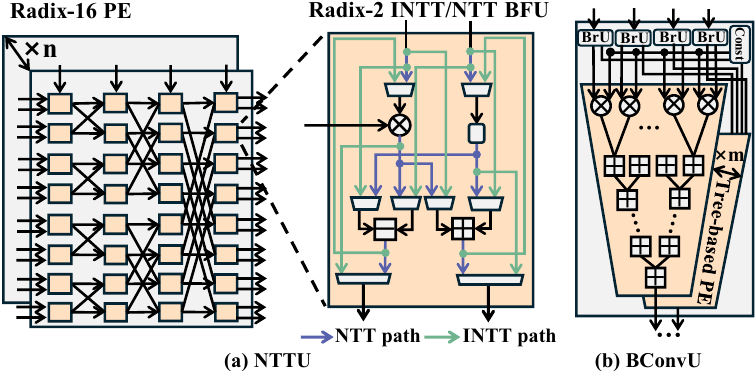}
    \caption{Microarchitectures of the iterative-based NTTU (a) and tree-based BConvU (b).}
    \label{fig:NTTU BConvU}
\end{figure}

Inter-operator overlap between (I)NTT and BConv is challenging because (I)NTT accesses different coefficients within one polynomial, whereas BConv processes coefficients from multiple polynomials concurrently, as depicted in Fig.~\ref{fig:dimension}(a). Existing designs~\cite{SHARP,ARK,F1,craterlake} offer high intra-polynomial but limited inter-polynomial parallelism in the NTTU, making its output throughput across limbs insufficient for the BConvU input and thus hindering effective inter-operator overlap. 

To balance the throughput of NTTU and BConvU, we adopt a configurable iterative radix-2 NTTU~\cite{scalableNTT} (Fig.~\ref{fig:NTTU BConvU}(a)). For each \textit{dnum} group, NTTUs are evenly distributed across all limbs required by BConv, ensuring parallel data supply and full pipelining. As BConvU typically needs fewer than fifteen limbs per decomposed group, NTTU parallelism remains sufficient. We further design a tree-based BConvU (Fig.~\ref{fig:NTTU BConvU}(b)), where each unit simultaneously receives one coefficient from all limbs within a decomposed group and performs pipelined tree reduction. NTTUs are dynamically shared between NTT and INTT to align with BConvU I/O demands. Through adaptive throughput matching and flexible scheduling, the xPU achieves effective INTT-BConv-NTT overlap (Fig.~\ref{fig:dimension}(d)), matching SHARP's performance on the IRF critical path while using lower NTT and BConv throughputs.

\subsection{Reduction of on-xPU Memory Capacity}\label{sec:on-xPU memory}
With the HERO optimization and the IRF mapping scheme (Sec.~\ref{sec:dataflow mapping}), all IPs are offloaded to the xMU, while the xPU primarily handles ModUps and ModDowns. Hence, the xPU memory mainly stores intermediate results of ModUp and ModDown. As noted in Sec.~\ref{sec:xPU}, the computation-communication pipeline streams ModUp outputs to the xMU as soon as they are produced, and ModDown starts upon receiving input, removing the need to buffer the full result. Therefore, only partial ModUp/ModDown data is cached, with the specific volume determined by the discrepancy between the throughput of xPU compute units and the off-chip transfer bandwidth. Simulations show that a 44 MB scratchpad can sustain fully pipelined ModUp and ModDown for IRF, and 84 MB is sufficient for the hybrid scheme, which necessitates one extra \textit{evk}. Moreover, both capacity configurations are sufficient to accommodate the baby-step (\textit{bs}) ciphertexts used in ciphertext polynomial evaluations based on the Peterson-Stockermeyer algorithm~\cite{improvedboot}, as well as the intermediate ciphertexts associated with ciphertext linear combinations offloaded to the xPU. These cases either occur at lower ciphertext levels, where the ciphertext size is relatively small, or during the bootstrapping polynomial computation, for which we adopt a low \textit{bs}. We term the two architectures HE\textsuperscript{2}-SM and HE\textsuperscript{2}-LM, both evaluated in Sec.~\ref{sec:end-to-end benchmarks}.

\subsection{Intra-Keyswtich Parallelism Exploitation on xPU}\label{sec:parallelism exploitation}
In ModUp/ModDown pipelines, two parallel execution paths exist:
\textbf{(1) critical path}: INTT$\rightarrow$BConv$\rightarrow$NTT, and 
\textbf{(2) secondary path}: direct preservation of original limbs without arithmetic operations.
The imbalance between the two stems from the heavier workload on NTT-domain ciphertexts, which are efficient for polynomial multiplication, whereas INTT-domain ciphertexts appear only before BConv.

\begin{figure}[]
    \centering
    \includegraphics[width=1\linewidth]{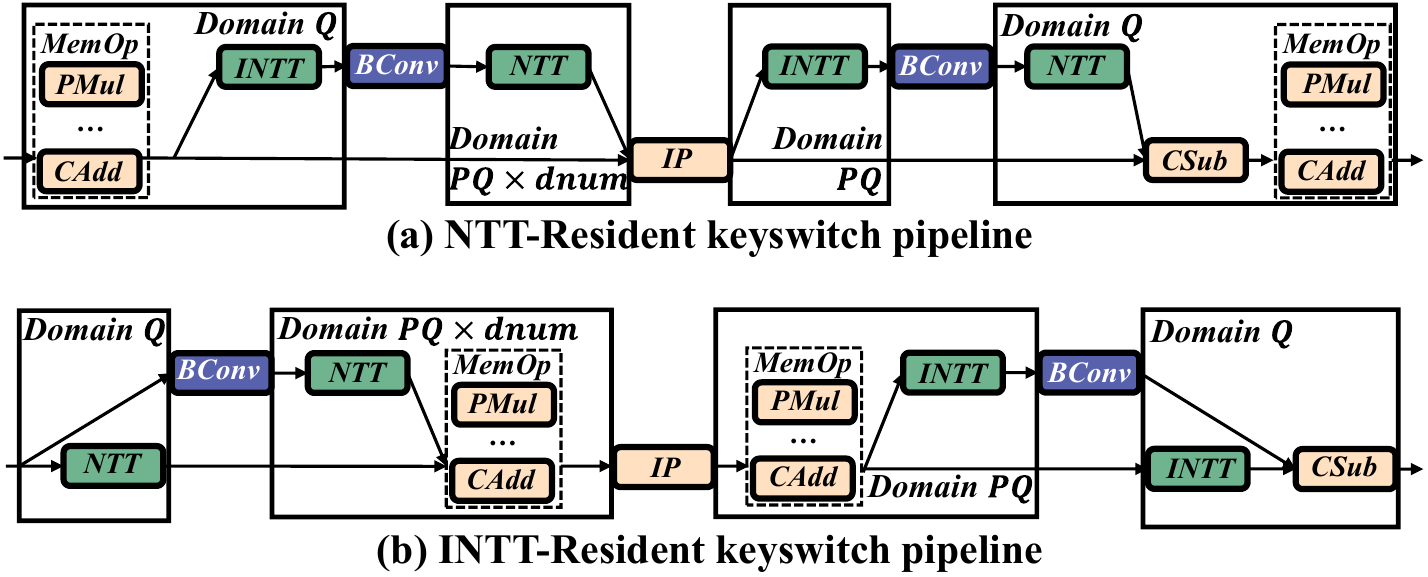}
    \caption{Trade-off of two types of keyswitches. (a) For NTT-domain ciphertexts, keyswitch operates with MemOps under the smaller modulus $Q$, but exhibits no parallelism between (I)NTT and BConv. (b) For INTT-domain ciphertexts, keyswitch executes MemOps at the larger modulus $PQ$ or $PQ\cdot\textit{dnum}$, yet achieves higher parallelism between (I)NTT and BConv.}
    \label{fig:keyswitch parallel}
\end{figure}

To exploit potential parallelism, we adopt an adaptive ciphertext-format management strategy on the xPU. Specifically, an \textbf{NTT-Resident} strategy is used for subgraphs involving PMul or CMul, while the \textbf{INTT-Resident} approach is applied to others. 
As shown in Fig.~\ref{fig:keyswitch parallel}, INTT-Resident strategy breaks INTT$\rightarrow$BConv$\rightarrow$NTT into parallel BConv$\rightarrow$NTT and NTT paths. Although this incurs extra MemOp overhead due to domain changes, the additional parallelism for the xPU and the xMU's high near-memory bandwidth can offset this cost, as illustrated in Fig.~\ref{fig:dimension}(e).

Moreover, an NTTU allocator is introduced on the xPU to dynamically balance the workload between the two parallel paths in the INTT‑Resident pipeline as the ciphertext level changes.

\subsection{Low-Overhead xMU Microarchitecture}\label{sec:xMU}

Our xMU is designed with a hardware-overhead-driven philosophy, minimizing added area and power during logic integration into HBM. After HERO optimization, only lightweight operations--CtAdd, PtMul, IP, and Autom--are mapped to the xMU. 
As shown in Fig.~\ref{fig:overview}(b) and (c), we adopt bank-level PE integration (following~\cite{dram_scarce_compute,1ynm,bank-pim3}) to maximize near-memory bandwidth while controlling power. Each xMU PE fetches 256-bit data from the global row buffer into a local buffer to hide bank access latency. 

Furthermore, since the operators offloaded to the xMU are uniformly vectorized across polynomials, we employ a row-major data layout that distributes each polynomial across all banks, allowing every PE to locally access its operands. To further reduce in-memory data movement, we propose MemOp fusion to eliminate the row-switch overhead for intermediate results during sequential MemOps such as IP and PMul shown in Fig.~\ref{fig:overview}(d).

In addition, the automorphism is implemented entirely within DRAM without extra hardware. By leveraging the shared multi-level controllers and local buffers--an approach inspired by~\cite{figaro}--data movement is performed hierarchically: via the global row buffer (2048 coeff/cycle) for intra-bank transfers, the bank I/O controller (128 coeff/cycle) for inter-bank transfers, and the GBus controller (32 coeff/cycle) for inter-bank-group transfers. This reuse of native DRAM data paths eliminates the need for new logic while achieving a 1.10$\times$ speedup over the two-level automorphism in F1~\cite{F1}, making it an efficient yet low-overhead complement to the xMU design.

Our xMU adopts HBM with 8 GB capacity and 1 TB/s bandwidth to balance computation and communication demands. This configuration ensures sufficient throughput for heterogeneous data exchange and supports the PKB fusion in HERO. Further sensitivity analysis is presented in Sec.~\ref{sec:sensitivity study}.

\begin{table}[]
\fontsize{12}{16}\selectfont
\centering
\caption{Modeling Configurations of HE\textsuperscript{2} and prior works. }
\label{tab:modeling}
\resizebox{1\linewidth}{!}{
\begin{threeparttable}
\begin{tabular}{c|c|c|c|c|cc}
\toprule
\multirow{2}{*}{\textbf{Configs}} & \multirow{2}{*}{\textbf{SHARP}} & \multirow{2}{*}{\textbf{FAST}} & \multirow{2}{*}{\textbf{FHENDI}} & \multirow{2}{*}{\makecell{\textbf{Anaheim}\\ \textbf{xMU}}} & \multicolumn{2}{c}{\textbf{HE\textsuperscript{2}-SM/LM}} \\
& & & & & \textbf{xPU} & \textbf{xMU} \\ \hline
\makecell{Word Width} & 36-bits & 36/60-bits & 46/51-bits & 28-bits & \multicolumn{2}{c}{36-bits} \\
\makecell{Core Freq.} & 1 GHz & 1 GHz & - & 0.38 GHz & 1 GHz & 0.45 GHz \\
\makecell{NTTU Throu. (w/ns)}\tnote{$\ast$} & 1024 & \makecell{1024/512} & 2048 & - & 768 & - \\
\makecell{BConvU Throu. (w/ns)} & 16384 & \makecell{16384/8192} & - & - & 672 & - \\
\makecell{EWEU Throu. (w/ns)}\tnote{$\S$} & 2048 & \makecell{2048/1024}& 1024 & 7760 & 512 & 5461 \\
\makecell{HBM BW (TB/s)} & 1 & 1 & 1 & 1.76 & \multicolumn{2}{c}{1} \\
\makecell{On-chip BW (TB/s)}\tnote{$\dagger$} & 36+36 & 72+72 & - & 140 & 36 & 96 \\
\makecell{On-chip Cap (MB)}\tnote{$\ddagger$} & 180+18 & 281 & - & - & 44/84 & - \\
Optimization for PKB & Min-KS & Hybrid & Hoisting & Hoisting & \multicolumn{2}{c}{Hoisting/Hybrid} \\
Dataflow Mode & EVF & EVF & EVF & IRF & \multicolumn{2}{c}{IRF/Hybrid} \\
\hline
Our FHE parameters & \multicolumn{6}{c}{\large $N=2^{16},L=35,L_{\text{eff}}=8,k=12,\alpha=12,\text{dnum}=3,\lambda=128\text{-bit}$} \\ \bottomrule
\end{tabular}
\begin{tablenotes}   
\item[$\ast$] Since NTTUs in HE\textsuperscript{2} are dynamically allocated between NTT and INTT, the throughput is averaged over all NTT and INTT in the keyswitch with different levels. Throughput is measured in words per nanosecond (w/ns). 
\item[$\S$] The throughput of EWEU is defined as that of IPs.
\item[$\dagger$] SRAM/register bandwidth for ASICs or xPUs, and near-memory bandwidth for xMUs.
\item[$\ddagger$] SRAM/register capacity for ASICs or xPUs.
\end{tablenotes}
\end{threeparttable}}
\end{table}

\begin{table}[]
    \centering
    \caption{Area and Peak Power Breakdown. }
    \label{tab:area and power}
    \resizebox{0.75\linewidth}{!}{
    \begin{tabular}{l|cc}
        \toprule
        \textbf{Modules}      & \textbf{Area (mm\textsuperscript{2})} & \textbf{Power (W)} \\\hline
        $96\times$NTTU  & 2.05 & 8.71 \\
        $672\times$BConvU     & 5.32  & 22.6 \\
        OF-Twist   & 0.12  & 0.53 \\
        EWEU       & 0.67  & 2.84 \\
        Scratchpad-SM (44 MB) & 9.60  & 7.90 \\ 
        Scratchpad-LM (84 MB) & 17.9 & 13.1 \\
        NoC        & 0.01  & 0.03 \\
        HBM PHY    & 29.6  & 31.9 \\
        \textbf{xPU-SM (7 nm)} & \textbf{47.4}  & \textbf{74.5} \\ 
        \textbf{xPU-LM (7 nm)} & \textbf{55.7}  & \textbf{79.7} \\ \hline
        Comp. Units  & 11.9 & 11.0 \\
        Registers    & 0.40 & 0.75 \\
        \textbf{Total xMU (12 nm)} & \textbf{12.2} & \textbf{11.8} \\ \hline
        \textbf{HE\textsuperscript{2}-SM (1 xPU 2 xMU)}  & \textbf{71.9} & \textbf{98.0} \\
        \textbf{HE\textsuperscript{2}-LM (1 xPU 2 xMU)}  & \textbf{80.2} & \textbf{103} \\
        BTS (7 nm)     & 374 & 163 \\
        CLake (7 nm)   & 223 & 320 \\
        ARK (7 nm)      & 418 & 281 \\
        SHARP (7 nm)    & 179 & - \\
        UFC (7 nm)      & 198 & - \\
        FAST (7 nm)     & 284 & 338 \\
        FHENDI (12 nm) & 890 & 628 \\
        \bottomrule
    \end{tabular}}
\end{table}

\section{Methodology}
\subsection{HE\textsuperscript{2} Evaluation}\label{sec:he2 evaluation}
We choose SHARP~\cite{SHARP} as the core baseline to evaluate different algorithms and architectural variants, since SHARP achieves high performance under low hardware overhead through short word-length design and efficient on-chip memory management. SHARP has demonstrated state-of-the-art energy-delay-area product (EDAP) in existing ASIC implementations~\cite{ARK,FAST}, which aligns with our motivation to investigate heterogeneous architectures that balance computational efficiency and hardware cost.

\noindent \textit{\textbf{Simulator.}} We develop cycle-accurate performance simulators to evaluate:
\textbf{(1) SHARP} by strictly following its technical specifications, replicating its architecture and Min-KS-based dataflow strategy. In our end-to-end benchmarks, our simulated performance differs from the results reported in the original paper by 1.20\% on average. We further model SHARP with hoisting by prefetching the required \textit{evks} before each hoisted PKB, thereby providing a convincing approximation of SHARP’s actual performance with hoisting. \textbf{(2) HE\textsuperscript{2}-LM}, with a larger 84 MB on-chip memory supporting hybrid dataflow. \textbf{(3) HE\textsuperscript{2}-SM}, a smaller variant with 44 MB on-chip memory, sufficient for ciphertexts but unable to store any \textit{evk}, thus only supporting the IRF. \textbf{(4) SHARP-xMU}, a heterogeneous hardware combining SHARP with NMP-enabled HBM, adopting IRF dataflow.

\begin{table*}[]
    \centering
    \begin{threeparttable}
    \caption{End-to-end Latency (\textnormal{ms}), EDP (\textnormal{J$\cdot$ms}), EDAP (\textnormal{J$\cdot$ms$\cdot$mm\textsuperscript{2}}) Evaluation. }
    \label{tab:end-to-end benchmarks}
    \begin{tabular}{l|ccc|ccc|ccc|ccc}
        \toprule
        & \multicolumn{3}{c|}{\textbf{Bootstrapping~\cite{improvedboot}}} & \multicolumn{3}{c|}{\textbf{HELR~\cite{HELR}}} & \multicolumn{3}{c|}{\textbf{ResNet-20~\cite{HEResNet}}} & \multicolumn{3}{c}{\textbf{ResNet-56~\cite{HEResNet}}} \\
        & Latency & EDP & EDAP & Latency  & EDP & EDAP & Latency & EDP & EDAP & Latency & EDP & EDAP \\ \hline
        Anaheim (GPU-NMP) & 29.3 & - & - & 41.2 & - & - & 1020 & - & - & 3476 & - & - \\
        BTS (ASIC) & 22.9 & - & - & 28.4 & - & - & 1910 & - & - & 6509 & - & - \\
        CLake (ASIC) & 6.32 & 9.91 & 4.68K & 15.2 & - & - & 321 & 28.7K & 13.6M & 1094 & 333K & 158M \\
        ARK (ASIC) & 3.52 & 1.67 & 699 & 7.42 & 5.54 & 2.32K & 125 & 2.03K & 848K & 426 & 23.6K & 9.85M \\
        SHARP (ASIC) & 3.12 & 0.94 & 168 & 2.53 & 2.56 & 458 & 99 & 648 & 116K & 337 & 7.51K & 1.34M \\
        UFC (ASIC) & 2.60 & 0.45 & 89.8 & 2.11 & 1.15 & 229 & 90 & 331 & 65.5K & 302 & 3.73K & 738K \\
        FAST (ASIC) & 1.38 & 0.20 & 56.8 & 1.33 & 2.20 & 625 & 61 & 595 & 169K & 205 & 6.72K & 1.91M \\
        FHENDI (NMP) & 1.56 & - & - & - & - & - & 83 & - & - & 284 & - & - \\
        \hline
        
        \textbf{HE\textsuperscript{2}-SM} & \textbf{1.42} & \textbf{0.16} & \textbf{11.2} & \textbf{1.79} & \textbf{0.87} & \textbf{62.5} & \textbf{69.7} & \textbf{234} & \textbf{16.8K} & \textbf{232} & \textbf{2.60K} & \textbf{186K} \\
        \textbf{HE\textsuperscript{2}-LM} & \textbf{1.33} & \textbf{0.13} & \textbf{10.7} & \textbf{1.70} & \textbf{0.75} & \textbf{59.9} & \textbf{71.9} & \textbf{219} & \textbf{17.5K} & \textbf{240} & \textbf{2.43K} & \textbf{194K} \\
        \bottomrule
    \end{tabular}
    \begin{tablenotes}   
        \item[$\ast$] ResNet‑56 performance of the existing designs is obtained by scaling the reported results on ResNet‑20 according to computation load.  
    \end{tablenotes}
    \end{threeparttable}
\end{table*}
\noindent \textit{\textbf{Hardware and Algorithmic Specifications.}} 
A comparison of the modeling configurations between HE\textsuperscript{2} and prior works is presented in Table~\ref{tab:modeling}. Compared with ASIC-based designs~\cite{SHARP,FAST}, HE\textsuperscript{2} xPU features lower module parallelism and hardware overhead yet delivers superior performance under limited resources. This improvement stems from HERO (Sec.~\ref{sec:HERO}), which exploits the hoisting potential to reduce ComOp burden for xPU, and the pipelined xPU design (Sec.~\ref{sec:xPU}), which enables inter-operator overlap. Compared with NMP-based works~\cite{fhendi}, we adopt a lightweight near-memory integration: only simple SIMD-style MemOps are offloaded to the xMU, and a fusion strategy (Sec.~\ref{sec:xMU}) ensures high near-memory bandwidth utilization similar to~\cite{flexmem} but with minimal hardware overhead.
We adopt the same algorithmic parameters as SHARP, as shown in Table~\ref{tab:modeling}. We use an FFT-like bootstrapping~\cite{improvedboot} with three stages. Min-KS~\cite{ARK} is employed to reduce \textit{evk} storage when neither hoisting nor HERO is applied. 

\noindent \textit{\textbf{BSGS Configurations.}}
In our evaluations of bootstrapping, HELR, and ResNet, we disable BSGS in both C2S and S2C under the 8 GB HBM capacity constraint. For BERT, BSGS is still adopted in the first FFT stage of C2S during its bootstrapping, with a baby step of 2 and a giant step of 32. This is because the first FFT stage operates at a relatively high ciphertext level. If BSGS is disabled in this stage, the resulting \textit{evk} would surpass the available HBM capacity. 

\subsection{Benchmarks}\label{sec:benchmarks}
\textbf{Bootstrapping} follows state-of-the-art fully-packed implementation~\cite{SHARP,ARK} with 8 efficient levels. \textbf{HELR}~\cite{HELR} is an ML workload which trains a binary classifier. We use a batch size of 1024 and report the average latency over 32 iterations. We evaluate \textbf{ResNet-20} and \textbf{ResNet-56}~\cite{HEResNet}, which are CNN models adopting a multiplexed packing method, with encrypted images of size 32$\times$32$\times$3 and batch size of 1. Furthermore, we evaluate a 12-layer \textbf{BERT-based} model~\cite{BERT} for a single inference with a 128$\times$768 input sequence. 

\section{Evaluation}\label{sec:evaluation}
\subsection{Implementation of HE\textsuperscript{2}}

We implement HE\textsuperscript{2} in RTL and estimate its power and area. We employ the 7 nm TSMC PDK for the xPU and the TSMC 12 nm PDK for the xMU PEs. We integrate the OF-Twist proposed in ARK~\cite{ARK} for twiddle factor generation, and implement the KSKGen proposed in~\cite{craterlake} for \textit{evk} generation in xMU PEs. The EWEU is implemented following SHARP~\cite{SHARP}, with four modular multipliers and two modular adders per unit, but only with one-quarter of SHARP's parallelism, dedicated to EWOs on the xPU. On-chip wiring and scratchpad memory are modeled using CACTI-6.0~\cite{CACTI}. Our xMU PEs are integrated within two HBM2 stacks~\cite{hbm,hbm2}, aligning with the off-chip memory in prior ASIC-based designs~\cite{ARK,SHARP,Trinity}, which provides a total of 1 TB/s off-chip bandwidth, and 8 GB capacity. The xPU operates at 1 GHz, while PEs within the xMU run at 450 MHz. 

Table~\ref{tab:area and power} shows the area and power comparison with existing works~\cite{BTS,craterlake,ARK,SHARP,fhendi}. On the one hand, by reducing both the large on-chip memory and per-operator parallelism, we significantly lower the area and power overheads of the xPU. On the other hand, the xMU PEs adopt a feasible bank-level design consistent with prior near-memory architectures~\cite{dram_scarce_compute,1ynm,bank-pim3}, maintaining peak power within the HBM's all-bank-interleave access budget~\cite{jedec,attacc} and operating within the \SI{85}{\celsius} thermal envelope. Our xMU area evaluation considers the logical integration density scaling factor in DRAM technologies according to~\cite{dram_logic_1.5_less_dense,dram_logic_2_less_dense}. The xMU PEs occupy only 11.1\% of the HBM module area, satisfy RTL-verified timing constraints, and preserve HBM bank I/O compatibility under standard design rules~\cite{jedec}, thereby demonstrating the feasibility for implementation~\cite{hbmpower,jedec}. 

\subsection{End-to-end Evaluation and Ablation Study}\label{sec:end-to-end benchmarks}
We evaluate HE\textsuperscript{2} on four benchmarks and compare the performance against a GPU-NMP heterogeneous design (Anaheim~\cite{Anaheim}), ASIC (BTS~\cite{BTS}, CLake~\cite{craterlake}, ARK~\cite{ARK}, SHARP~\cite{SHARP}, UFC~\cite{UFC}, and FAST~\cite{FAST}), and NMP (FHENDI~\cite{fhendi}) accelerators, as shown in the Table~\ref{tab:end-to-end benchmarks}. 
Overall, HE\textsuperscript{2} achieves the optimal EDP and EDAP, and state‑of‑the‑art latency performance. \textbf{(1)} Compared with Anaheim~\cite{Anaheim}, which is a GPU-NMP heterogeneous design applying hoisting on raw CKKS programs, HE\textsuperscript{2} achieves an average 22.0$\times$ improvement in latency. This is because: \textit{a)} Executing FHE operators on a general-purpose GPU architecture makes it difficult to achieve efficient overlap between operators. \textit{b)} The lack of a customized keyswitch dataflow design in Anaheim prevents computation communication overlap, failing to address the costly communication on the critical path. \textit{c)} HERO fuses PKBs in the program rather than directly applying hoisting, achieving much higher keyswitch parallelism than the original CKKS program used in Anaheim. \textbf{(2)} FAST~\cite{FAST} is the fastest prior ASIC accelerator, leveraging an advanced keyswitch algorithm~\cite{klss} at the cost of considerable compute resources and on-chip storage. As shown in Table~\ref{tab:area and power}, FAST's area and peak power are 3.54$\times$ and 3.28$\times$ those of HE\textsuperscript{2}-LM, respectively. Thus, HE\textsuperscript{2} delivers 2.49$\times$ and 8.81$\times$ improvements in EDP and EDAP over FAST, while achieving comparable latency. In addition, \textbf{(3)} compared to the state‑of‑the‑art NMP accelerator FHENDI~\cite{fhendi}, HE\textsuperscript{2} demonstrates similar end‑to‑end performance with 11.1$\times$ reduction in area and 6.10$\times$ reduction in peak power. Overall, our performance improvement can be attributed to two factors: \textit{\textbf{(1) the co-design of DFG optimizations and heterogeneous dataflow}}, and \textit{\textbf{(2) architectural optimizations.}} 

\begin{figure*}[]
    \centering
    \includegraphics[width=1\linewidth]{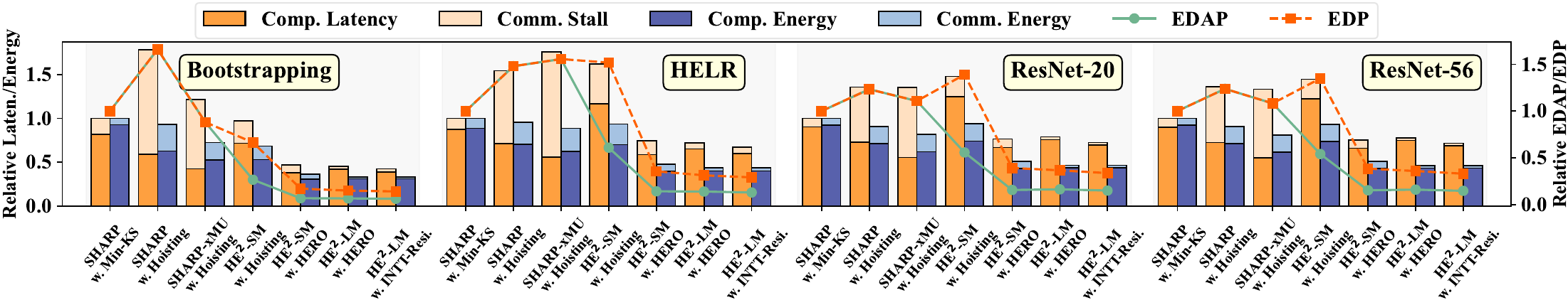}
    \caption{Ablation study results. SHARP w. Hoisting exhibits inadaptability to hoisting due to long memory stalls (2nd column). SHARP‑xMU is a heterogeneous design adopting IRF dataflow, with its xPU aligned with SHARP, which shows partial hoisting compatibility (3rd column). HE\textsuperscript{2}-SM, using the same IRF dataflow, achieves similar performance to SHARP-xMU but with fewer stalls (4th column), as its dual-pipelined xPU hides communication. And HERO (5th column) shows significant computation and communication reduction compared to hoisting. HE\textsuperscript{2}-LM adopts a hybrid dataflow (6th column), further mitigating communication where hoisting is inapplicable. The INTT-Resident (7th column) strategy enhances xPU computation parallelism.}
    \label{fig:ablation}
\end{figure*}

\begin{figure}[]
    \centering
    \includegraphics[width=1\linewidth]{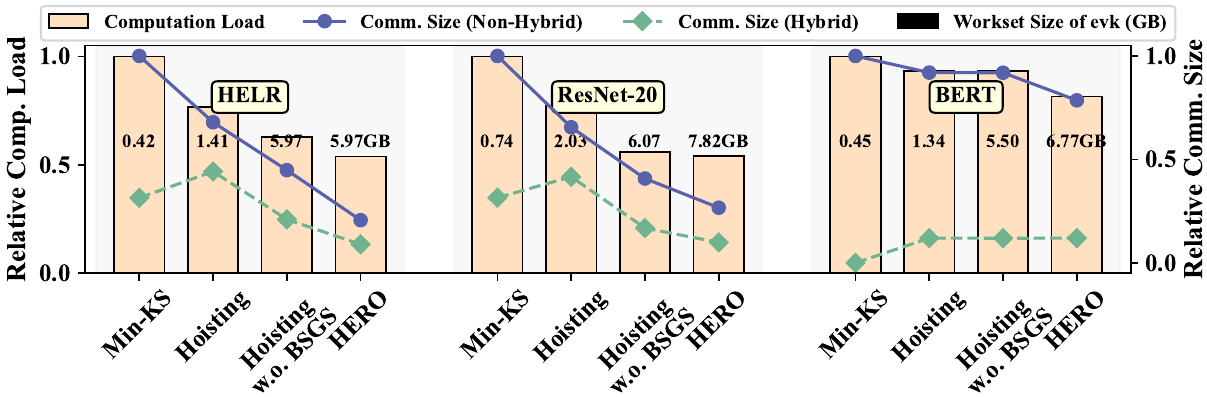}
    \caption{Comparison of algorithmic optimizations effects on HE\textsuperscript{2}. HERO enhances computational and communication efficiency at the cost of increased \textit{evk} storage. Compared with applying hoisting in the original program or in the program with BSGS disabled (Hoisting w.o. BSGS), HERO achieves a greater reduction in both computation and communication. }
    \label{fig:DFG opt impact}
\end{figure}
To provide deeper insights, we carry out an ablation study based on the SHARP baseline, which is selected for its state-of-the-art EDAP and hardware efficiency, as shown in Fig.~\ref{fig:ablation}. For SHARP with the EVF dataflow (1st and 2nd column), we apply two algorithmic optimizations, Min-KS and hoisting. The results show that the low \textit{evk} reuse introduced by hoisting increases off-chip memory access under EVF, causing performance degradation compared with Min-KS. 

In contrast, a heterogeneous architecture with IRF dataflow (3rd column) shifts from reusing \textit{evk} on the xPU to reusing intermediate results across the \textit{evks} stored on the xMU, thereby partially reducing the communication overhead. However, low parallelism PKBs gain little from hoisting. In such cases, IRF leads to intermediate ciphertext movement that surpasses the cost of \textit{evk} loading, and communication on the critical path, resulting in significant stalls (e.g., the 3rd column of HELR).

To address communication stalls and provide more flexible dataflow support, \textbf{first}, the HE\textsuperscript{2} dual-level overlapping xPU design hides most communication latency while maintaining performance (4th column); \textbf{second}, increased PKB parallelism in HERO amplifies hoisting benefits for both computation and communication (5th column); \textbf{third}, hybridizing IRF and EVF with an additional on-xPU buffer sufficient for one \textit{evk} further reduces communication stalls (6th column). Specifically, for PKBs with parallel keyswitches, hoisting and IRF reuse intermediate results to minimize data movement, whereas for PKBs with a single keyswitch, EVF preloads the required \textit{evk} on xPU, since the overhead of preloading one \textit{evk} is smaller than that of moving intermediate results. \textbf{Finally}, the INTT-Resident mechanism enhances xPU's computation parallelism (7th column). In the hybrid dataflow, the IRF region is bounded by communication, while the EVF region is limited by computation latency. Therefore, the INTT-Resident optimization mainly benefits the EVF parts. 
In summary, HE\textsuperscript{2}-LM achieves a 9.23$\times$ EDAP and 4.13$\times$ EDP improvement over SHARP. Moreover, communication stalls account for only 6.67\% of the total execution time, and communication energy overhead is reduced to 6.60\%. 

\subsection{Impact of HERO on Computation and Communication}\label{sec:impact of DFG optimization}
We evaluate the algorithmic optimizations of HERO.
As shown in Fig.~\ref{fig:DFG opt impact}, hoisting merges ModUps and ModDowns among parallel keyswitches to cut communication and computation (2nd column). When BSGS is disabled (3rd column), or when BERT adopts a large gap between the baby‑step and giant‑step parameters (see Sec.~\ref{sec:he2 evaluation} for details), more parallel keyswitches occur in bootstrapping, yielding further gains. Finally, HERO consolidates the low‑parallelism PKBs across the program (4th column), achieving the optimal overall performance.

\begin{figure}[]
    \centering
    \includegraphics[width=1\linewidth]{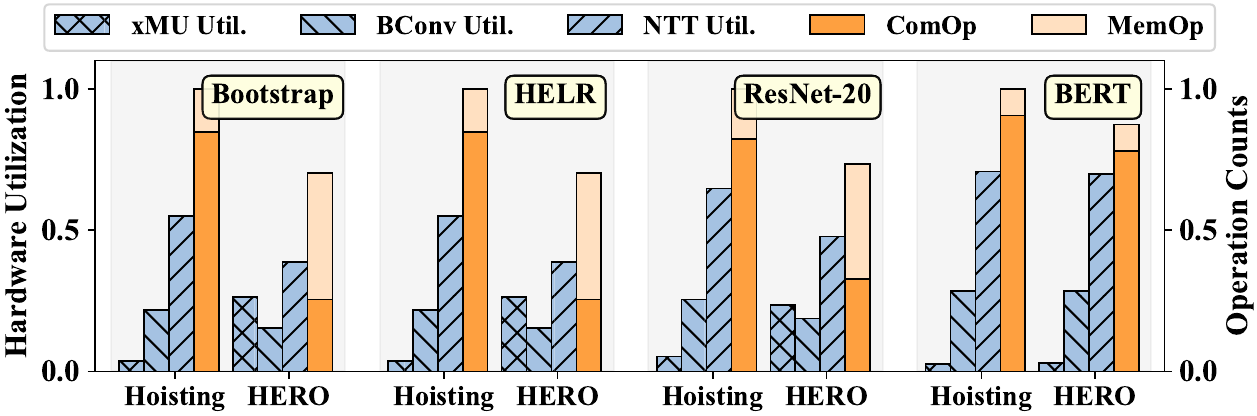}
    \caption{Hardware utilization and operation counts of HE\textsuperscript{2}. HERO increases the proportion of MemOps and the utilization of xMU. Both hoisting and HERO are evaluated under an IRF dataflow.}
    \label{fig:utilization}
\end{figure}

Furthermore, the dataflow strategy strongly influences the communication volume. IRF markedly reduces communication for PKBs with high parallelism, whereas for PKBs with low parallelism, the communication cost of loading a single \textit{evk} is smaller than that of transferring intermediate results in IRF, making EVF more efficient. As a result, the hybrid strategy, which selectively applies IRF and EVF, achieves lower communication overhead than pure IRF.

\subsection{Hardware Utilization Analysis}
We evaluate the hardware utilization, as shown in Fig.~\ref{fig:utilization}. When HE\textsuperscript{2} adopts the IRF dataflow with hoisting (1st column), the overall execution becomes bounded by xPU computation rather than memory, in contrast to SHARP~\cite{SHARP}, where hoisting exposes memory bottlenecks. This shift occurs because IRF eliminates the need to load \textit{evks}. With HERO enabled (2nd column), the range and efficiency of hoisting are further enhanced, leading to a higher proportion of MemOp workloads and increased xMU utilization.  

\subsection{Sensitivity}\label{sec:sensitivity study}
\begin{figure}[]
    \centering
    \includegraphics[width=1\linewidth]{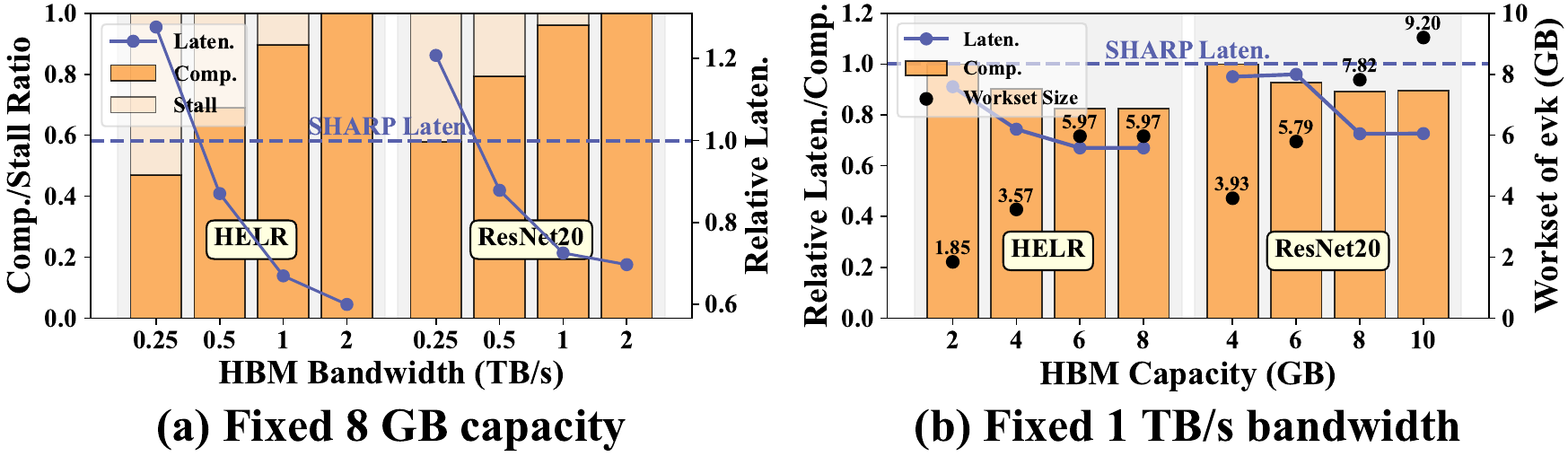}
    \caption{Performance scaling of HE\textsuperscript{2} with respect to HBM bandwidth and capacity shows a gradual saturation trend, demonstrating the effectiveness of communication optimization in mitigating bandwidth requirements. And the optimal PKB fusion solution found by HERO can be achieved within an 8 GB memory budget. SHARP adopts an HBM with 1 TB/s bandwidth and 8 GB capacity.}
    \label{fig:sensitivity}
\end{figure}
We investigate HE\textsuperscript{2}'s sensitivity to variations in the NMP-enabled HBM (xMU) bandwidth and capacity. As shown in Fig.~\ref{fig:sensitivity}(a), increasing the bandwidth reduces communication stalls, leading to performance improvement. Results show that \textbf{\textit{HE\textsuperscript{2} outperforms SHARP at the same 1 TB/s bandwidth and maintains comparable performance to SHARP with bandwidth at least 0.5 TB/s.}}

The experiment shown in Fig.~\ref{fig:sensitivity}(b) evaluates the effectiveness of HERO’s PKB fusion strategy under different capacity constraints. With the same capacity of 8 GB, HE\textsuperscript{2} achieves higher performance than SHARP. Furthermore, \textbf{\textit{8 GB is sufficient to identify the optimal fusion strategies}} for the benchmarks tested. Although a larger memory capacity enables more aggressive fusion strategies, the PKBs at this point already exhibit substantial parallelism. Further fusion results in an explosive increase in IP computation, which in turn degrades performance.

\section{Related work}
\noindent\textbf{ASIC- and FPGA-based FHE accelerators.} 
HEAX~\cite{heax} and Roy et al.~\cite{anotherfpga} proposed FPGA accelerators for CKKS, but without support for the bootstrapping. Poseidon~\cite{Poseidon} and FAB~\cite{FAB} further enable arbitrary-depth CKKS computation with bootstrapping, but their acceleration performance remains lower compared to ASIC implementations. The first CKKS accelerator F1~\cite{F1} only supports shallow benchmarks. Craterlake~\cite{craterlake}, BTS~\cite{BTS}, and ARK~\cite{ARK} designed highly parallel modules combined with large on-chip memory to support deep CKKS applications and achieve high acceleration performance. SHARP~\cite{SHARP} further identified 36-bit datapaths as the most efficient choice, which reduces the hardware overhead. Alchemist~\cite{alchemist}, Trinity~\cite{Trinity}, and UFC~\cite{UFC} further proposed more general-purpose designs that support both CKKS and TFHE schemes. FAST~\cite{FAST} employs a more advanced keyswitch algorithm, which significantly reduces computation cost and enhances acceleration performance.

\noindent\textbf{NMP-based FHE accelerators.} 
Gupta et al.~\cite{UPMEM} proposed a DRAM-based NMP solution based on UPMEM~\cite{real_upmem} for BFV scheme acceleration. MemFHE~\cite{MemFHE} and FHE-PIM~\cite{FHE-PIM} leverage an RRAM-based NMP architecture for FHEW scheme acceleration. These designs cannot be extended to CKKS with more complex primitives. FHENDI~\cite{fhendi} adopted an HBM-based design, where massively parallel NMP units and parallel bootstrapping enable high performance, but its large area and power overhead limit manufacturability. FlexMem~\cite{flexmem} revealed the low near‑memory bandwidth utilization but used a complex in‑memory network restricted by the available metal layers in DRAM technology~\cite{ndpbridge}.

\noindent\textbf{Hoisting-based works.} Anaheim~\cite{Anaheim} applied hoisting to the raw CKKS program and identified the bottleneck in MemOps, and proposed a heterogeneous GPU-NMP architecture. However, its hoisting effectiveness is limited by low keyswitch parallelism in the native program. Moreover, without a customized xPU design or dataflow optimization for computation-communication balance, the overall acceleration remains limited. FAST~\cite{FAST} applied hoisting at low ciphertext levels without restructuring the program to better leverage it. Orion\cite{orion} represented convolutions as plaintext-ciphertext matrix multiplications to exploit BSGS and hoisting. Nevertheless, Orion ignored the effect of BSGS on the efficiency of hoisting and exhibited lower generality compared with HERO.

\section{Conclusion}
We present the first ASIC-NMP heterogeneous CKKS accelerator \textbf{HE\textsuperscript{2}}, which, for the first time, addresses the long-standing dilemma between high hardware cost and acceleration performance commonly encountered in both ASIC and NMP monolithic designs. 
To tackle the xPU-xMU communication bottleneck caused by the complex data dependencies between operators, we introduce algorithmic optimizations and microarchitectural enhancements, which effectively reduce communication frequency and hide latency. 
HE\textsuperscript{2} achieves 1.66$\times$ and 1.17$\times$ speedups over state-of-the-art ASIC and NMP accelerators, as well as 2.49$\times$ and 12.4$\times$ reductions in area, thereby highlighting a practical design avenue for future FHE accelerators. 

\section*{Acknowledgments}
We thank the anonymous reviewers and shepherd for their insightful comments and suggestions. This work is partially supported by the NSF of China (Grants No.62341411, 62222214), Strategic Priority Research Program of the Chinese Academy of Sciences (Grants No.XDB0660200, XDB0660201, XDB0660202), and Youth Innovation Promotion Association CAS.

\bibliographystyle{IEEEtranS}
\bibliography{refs}

\end{document}